\documentclass[%
reprint,
superscriptaddress,
showpacs,preprintnumbers,
amsmath,amssymb,
aps,
pre,
]{revtex4-2}

\usepackage{color,mathtools}
\usepackage[pdftex]{graphicx} 
\usepackage{dcolumn}
\usepackage{bm}
\usepackage{textcomp} 

\begin {document}

\title{Generalized Langevin dynamics for single beads in linear elastic network
}


\author{Soya Shinkai}%
\affiliation{Laboratory for Developmental Dynamics, RIKEN Center for Biosystems
  Dynamics Research, Kobe 650-0047, Japan }%

\author{Shuichi Onami}%
\affiliation{Laboratory for Developmental Dynamics, RIKEN Center for Biosystems
  Dynamics Research, Kobe 650-0047, Japan }%

\author{Tomoshige Miyaguchi}
\email{mygch@wakayama-u.ac.jp}
\affiliation{%
  Department of Systems Engineering, Wakayama University, 930 Sakaedani, Wakayama 640-8510, Japan}


%
%


\date{\today}


\begin{abstract}
  We derive generalized Langevin equations (GLEs) for single beads in linear
  elastic networks. In particular, the derivations of the GLEs are conducted
  without employing normal modes, resulting in two distinct representations in
  terms of resistance and mobility kernels.  The fluctuation-dissipation
  relations are also confirmed for both GLEs. Subsequently, we demonstrate that
  these two representations are interconnected via Laplace
  transforms. Furthermore, another GLE is derived by utilizing a projection
  operator method, and it is shown that the equation obtained through the
  projection scheme is consistent with the GLE with the resistance kernel.  As
  simple examples, the general theory is applied to the Rouse model and the ring
  polymer, where the GLEs with the resistance and mobility kernels are
  explicitly derived for arbitrary positions of the tagged bead in these models.
  Finally, the GLE with the mobility kernel is also derived for the elastic
  network with hydrodynamic interactions under the pre-averaging approximation.
  
\end{abstract}

\maketitle

\section {Introduction}
Not only do macromolecules such as proteins diffuse in solutions, but components
comprising the macromolecules, such as monomers and amino acids (referred to as
beads hereafter), also exhibit time-dependent fluctuations. To describe such
internal motion of macromolecules, elastic network models have been widely
employed \cite{tirion96, haliloglu97,bahar98, atilgan01,burioni04, reuveni10}. 
Beads in the elastic networks are connected by Hookean springs. The simplest
example of such models is the ideal Rouse chain \cite{rouse53, doi86}, but the
elastic network models also include polymers with network structures, in which a
bead can be linked to more than two beads. Polymers with such network
structures, forming branches and crosslinks, are also considered important for
modeling three-dimensional genome structures \cite{le18,liu19,shi19,shinkai20}.

Experimental studies have revealed complex properties in single beads dynamics,
and therefore, elucidating equations of motion that can describe such intricate
behaviors of single beads is essential for analyzing experimental data.
For instance, motions of single nucleosomes and single genomic loci within
chromatin fibers exhibit subdiffusion \cite{bronstein09, weber10, hihara12,
  shinkai16, ohishi22, nozaki23}. Furthermore, by indirectly monitoring distance
fluctuations between a pair of beads in a protein, it was reported that these
fluctuations can be described by a generalized Langevin equation (GLE)
\cite{kou04,min05}.

In theoretical studies, it was demonstrated that single beads dynamics of a
linear chain, such as the Rouse model, can be described by GLEs, as analytically
proven in Refs.~\cite{panja10, panja10b, lizana10, vandebroek17, sakaue13,
  saito15,tian22} by using normal modes. For instance, in Refs.~\cite{panja10,
  panja10b, vandebroek17}, it was shown that the middle bead of the Rouse and
other linear polymer models follows a GLE with a resistance memory
kernel. Meanwhile, in Refs.~\cite{sakaue13, saito15}, a tagged bead in the
linear chains was shown to follow a GLE with a mobility memory kernel,
regardless of the position of the tagged bead in the chain. However, a
derivation of the GLEs for bead motions in the elastic network model has not
been undertaken.


To the best of the authors' knowledge, the derivations of the GLEs for polymer
models have relied on normal mode techniques \cite{panja10, panja10b, sakaue13,
  vandebroek17}.  Contrastingly, in the projection operator formalism, a
slightly different form of the GLE is derived \cite{hansen90, dhont96} without
relying on the normal modes. Considering that the derivations of the GLEs in
linear polymer models can be viewed as a concrete example of the projection
operator formalism \cite{panja10b}, it is anticipated that the GLEs for the
polymer models and the elastic networks can be derived without resorting to the
normal modes.  The derivations of the GLEs in linear polymer models were carried
out by eliminating modes other than the tagged bead from the equations of motion
\cite{panja10, panja10b, sakaue13, vandebroek17}. However, it has remained
unclear how such procedures are related to the projection operator scheme.

In this paper, we show that single-bead motion of the linear elastic networks
can be described by the GLEs.
We utilize supervector notation \cite{dhont96}, which plays a crucial role in
the derivations of the GLEs. In a projection operator analysis, we employ the
Hermitian conjugate of the Smoluchowski operator \cite{dhont96}. After deriving
the GLEs, normal-mode equations of motion are constructed with careful attention
to preserving translational symmetry in space. Numerical simulations are carried
out to validate theoretical results for the Rouse model. Hydrodynamic
interactions (HIs) are incorporated into the model through a preaveraged
mobility tensor \cite{doi86}.

Firstly, we derive two GLEs for bead motion in linear elastic network models,
one with a resistance kernel and the other with a mobility kernel. In
particular, both GLEs are derived without resorting to normal modes. Secondly,
another GLE is derived using the projection operator method, and it is
demonstrated that this third GLE is consistent with the GLE with the resistance
kernel, although they are not equivalent. Since normal modes are not used in
these derivations, the correspondence between these GLEs becomes clear. Thirdly,
by using normal modes, we derive diagonalized GLEs with the resistance and
mobility kernels, and directly demonstrate that these GLEs are interconnected
via Laplace transforms.  Fourthly, the present framework is applied to the ideal
Rouse model, revealing that the memory kernels vary slightly depending on the
locations of the tagged bead in the Rouse chain. Finally, the GLE with the
mobility kernel is derived for the system with the HI.

This paper is organized as follows. In Sec.~\ref{s.model}, we introduce the
linear elastic network and the supervector notation. In
Sec.~\ref{s.derivation-gle}, the three GLEs are derived for single-bead motion
in the linear elastic network.  Additionally, the fluctuation-dissipation
relations (FDRs) are confirmed in this section. In Sec.~\ref{s.normal-modes},
the normal modes are introduced, and the memory kernels and correlated noises
are represented with these modes.  In Sec.~\ref{s.rouse-model}, the present
framework is applied to single-bead dynamics of simple polymer models: the Rouse
model and the ring polymer. In Sec.~\ref{s.HI}, for the linear elastic network
with the HI, the GLE with the mobility kernel is derived.  Section
\ref{s.discussion} is devoted to a discussion.

\section {Model}\label{s.model}

In this section, we introduce the linear elastic network, which is then
represented with the supervector notation. In addition, reduced vectors and
matrices are defined; they play a crucial role in derivations of the GLEs.


\begin{figure}[t!]
  \centerline{\includegraphics[width=8.5cm]{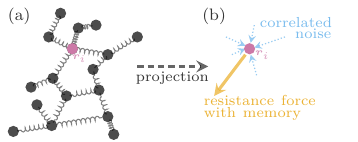}}
  \caption{\label{f.model} (a) Schematic illustration of the elastic network
    model. The circles and springs represent beads and harmonic interactions,
    respectively. These harmonic interactions are defined by the matrix $L$ in
    Eq.~(\ref{e.rGNM.svec}). (b) By projecting out the beads $\bm{r}_j$ other
    than the tagged one $\bm{r}_i$, we derive the GLE with the resistance kernel
    [Eq.~(\ref{e.gle.w.registance-kernel})]. In the GLE, the interactions with
    the other beads $\bm{r}_j$ are divided into two components: a resistance
    force with memory (solid arrow) and a correlated noise (dotted arrows).}
\end{figure}

\subsection {Linear elastic network}

We study a linear elastic network model with $N$ beads, illustrated in
Fig.~\ref{f.model}(a). This model is described by the Langevin equation
\cite{doi86, dhont96} as
\begin{equation}
  \label{e.rGNM.1}
  \gamma\frac {d \bm{r}_m(t)}{dt} 
  =
  \sum_{n=1}^{N} k_{mn} [\bm{r}_n(t) -\bm{r}_m(t)]
  +
  \bm{\xi}_m(t),
\end{equation}
with $m=1,\dots,N$. Here, $\bm{r}_m(t)$ is the position vector of the $m$th
bead, and $\gamma$ is a friction constant. Moreover, $\bm{\xi}_m(t)$ is a
three-dimensional Gaussian white noise; if the system is in equilibrium,
$\bm{\xi}_m(t)$ should satisfy the FDR
\begin{equation}
  \label{e.<xi-xi>}
  \left\langle \bm{\xi}_m(t_1) \bm{\xi}_n(t_2) \right\rangle
  =
  2k_BT \gamma \delta (t_1-t_2)\delta_{mn} I_3.
\end{equation}
Here $k_B$ is the Boltzmann constant, $T$ is the absolute temperature, $I_n$ is
the $n\times n$ identity tensor, $\delta(t)$ is the delta function, and
$\delta_{mn}$ is the Kronecker delta. In this paper, we neglect the excluded
volume effect, whereas the HI is incorporated in the subsection after next.

In addition, $k_{mn}$ in Eq.~(\ref{e.rGNM.1}) represents a spring constant
between the $m$th and $n$th beads; it is assumed that $k_{mn}$ is symmetric,
i.e., $k_{mn} = k_{nm}$, and that the diagonal entries are zero, $k_{nn}=0$.
For physical modeling of three-dimensional genome data, it is important to
accommodate the possibility of some spring constants being negative
\cite{shinkai20}. Consequently, instead of assuming $k_{mn}\geq 0$, we only
require that one eigenvalue of the Kirchhoff matrix $L$, which is defined below,
is zero, while the remaining $N-1$ eigenvalues are positive to ensure
thermodynamic stability. This allows for the possibility of some $k_{mn}$ being
negative. Thus, the present model is termed a linear elastic network with
non-negative eigenvalues.


We define an $N\times N$ matrix $L$ by $(L)_{mn} := l_{mn}$, where
$l_{mn} :=\delta_{mn} d_n - k_{mn}$ with $d_n := \sum_{m=1}^{N} k_{mn}$.  Note
that $L$ is symmetric, $l_{mn} = l_{nm}$.  Moreover, we employ supervector
notation $\bm{r} := (\bm{r}_1,\dots, \bm{r}_N)$ and
$\bm{\xi} := (\bm{\xi}_1,\dots, \bm{\xi}_N)$.  Then, Eq.~(\ref{e.rGNM.1}) is
rewritten as \cite{rouse53}
\begin{equation}
  \label{e.rGNM.svec}
  \gamma\frac {d \bm{r}(t)}{dt} 
  =
  - L \cdot \bm{r}(t)  +  \bm{\xi}(t),
\end{equation}
where the dot symbol represents the inner product of a (super)matrix and a
(super)vector. We also use the dot symbol to denote the inner product of two
(super)vectors, as in $\bm{v} \cdot \bm{w}$. In this notation, it is irrelevant
whether a vector is a column or a row vector [Exceptions are vectors used in
block matrices, such as in Eqs.~(\ref{e.L.vector_notation}), (\ref{e.JMJ}) and
(\ref{e.LML0}), where we distinguish between row and column vectors with a
transpose sign. See details below]. For example, the expression
$\bm{v}^t A\bm{w}$ in matrix notation can be simply written as
$\bm{v} \cdot A \cdot \bm{w}$.  In contrast, $\bm{v}\bm{w}$ represents the
tensor product of two vectors, resulting in a rank-two tensor.

An inner product (contraction) of two (super)matrices, say $A$ and $B$, is
denoted simply as $AB$. This notation is unambiguous, because tensor products of
matrices of the same size do not appear in this article.

With the supervector notation, the FDR in Eq.~(\ref{e.<xi-xi>}) is expressed as
\begin{equation}
  \label{e.<xi-xi>.svec}
  \left\langle \bm{\xi}(t_1) \bm{\xi}(t_2) \right\rangle
  =
  2k_BT \gamma \delta (t_1-t_2) I_3I_N,
\end{equation}
where $I_3I_N$ should be understood as an $N\times N$ matrix, with the $(i,j)$
element being $I_3\delta_{ij}$ (In other words, $I_3$ should be treated as if it
is a scalar). In some literature, $I_3I_N$ is denoted as $I_3\bigotimes I_N$
\cite{bahar98}, but here we employ the brief notation.

Note that the matrix elements $l_{mn}$ of $L$ satisfy
\begin{equation}
  \label{e.sum_l_mn=0}
  \sum_{n=1}^{N}l_{mn} = 0 \qquad (m=1,\dots,N),
\end{equation}
which is a manifestation of translational symmetry.  From
Eq.~(\ref{e.sum_l_mn=0}), it follows that $L$ has a zero eigenvalue with the
associated eigenvector $\bm{1}_N:=(1,\dots,1)$, i.e., $L \cdot \bm{1}_N = 0$. As
stated above, we assume that one eigenvalue of $L$ is zero and the remaining
$N-1$ eigenvalues are positive, i.e., $L$ is positive semidefinite.

\subsection {Reduced vector and matrix}\label{s.reduced-vec-mat}

We designate the $i$th bead as the tagged bead, and derive equations describing
its dynamics in the subsequent section. To accomplish this, we introduce an
operation, which is closely related to the projection operation. For an
$N$-dimensional vector $\bm{u} = (u_1,\dots,u_N)$ and an $N$-dimensional
supervector $\bm{w}=(\bm{w}_1,\dots, \bm{w}_N)$, we define $(N-1)$-dimensional
vector $\bm{u}'$ and supervector $\bm{w}'$ as
\begin{align}
\bm{u}' & = (u_1,\dots u_{i-1},u_{i+1},\dots,u_N), \\  
\bm{w}' & = (\bm{w}_1,\dots,\bm{w}_{i-1}, \bm{w}_{i+1},\dots, \bm{w}_N).   
\end{align}

Similarly, we define a reduced matrix $L'$ of order $N-1$ by eliminating the
$i$th row and column from $L$. More precisely, let us express the matrix $L$ as
$L = \left[\bm{l}_1, \dots, \bm{l}_N\right]$, where $\bm{l}_j$ is a column
vector corresponding to the $j$th column of $L$. Note also that we use the
square brackets to denote a matrix, thereby distinguishing it from a
supervector. Then $L'$ can be written by
\begin{equation}
  \label{e.L.vector_notation}
  L' = \left[\bm{l}'_1, \dots,\bm{l}'_{i-1}, \bm{l}'_{i+1}, \dots, \bm{l}'_N\right].
\end{equation}
In block matrices, such as in Eq.~(\ref{e.L.vector_notation}), vectors with a
transpose sign are always considered row vectors, while those without the sign
are considered column vectors [See Eqs.~(\ref{e.JMJ}) and (\ref{e.LML0})].  The
prime represents an operation of dimensionality reduction of vectors and
matrices.

It may be worth noting a connection to the projection operator formalism
here. In that formalism, a projection $\mathcal{P}$ onto a slow variable is
introduced, and $\mathcal{Q}$ is also defined as a complementary projection
satisfying $\mathcal{P} + \mathcal{Q}= \mathcal{I}$ with $\mathcal{I}$ being the
identity operator. As demonstrated in Sec.~\ref{s.projection}, the reduction of
$L$ to $L'$ is related to the action of $\mathcal{Q}$ to the Hermitian conjugate
of the Smoluchowski operator $\mathcal{L}_S^{\dagger}$ [See Eq.~(\ref{e.QLr'})].

By using these vectors and matrices, Eq.~(\ref{e.sum_l_mn=0}) can be rewritten
as $\sum_{n=1}^{N} \bm{l}'_n = 0$, or equivalently
\begin{equation}
  \label{e.ell_i}
\bm{l}_i' = -\sum_{n=1, n\neq i}^{N} \bm{l}'_n = - L'\cdot \bm{1}.
\end{equation}
Here, $\bm{1}$ represents an $N-1$ dimensional vector with all elements being
unity, i.e., $\bm{1} =: (1,\dots,1)$.  It is important to note that $L'$ is also
a symmetric matrix and is positive definite (See Appendix
\ref{s.positiv-definite}).

Utilizing the relation
$L\cdot \bm{r} = \bm{r}_1 \bm{l}_1 + \dots + \bm{r}_N \bm{l}_N$ and
Eq.~(\ref{e.ell_i}), we obtain
\begin{equation}
  \label{e.(Lr)'}
  (L \cdot \bm{r})'
  = L' \cdot \bm{r}' + \bm{r}_i \bm{l}_i'
  = L' \cdot (\bm{r}' - \bm{r}_i \bm{1}).
\end{equation}
Again, note that $\bm{r}_i \bm{l}_i'$ should be regarded as a supervector, each
element of which is given by $\bm{r}_il_{ji}$ , i.e., $\bm{r}_i$ should be
treated as if it is a scalar. Similarly, $\bm{r}_i \bm{1}$ is a supervector,
with each element being $\bm{r}_i$.
\subsection {Hydrodynamic interaction}
If we take into account the HI with pre-averaging approximation \cite{doi86},
the equation of motion [Eq.~(\ref{e.rGNM.svec})] is rewritten as
\begin{equation}
  \label{e.rGNM.svec.HI}
  \frac {d \bm{r}(t)}{dt} 
  =
  - L_H \cdot \bm{r}(t) +  \bm{\xi}_H(t),
\end{equation}
with $L_H := HL $ and $\bm{\xi}_H = H \cdot \bm{\xi}$, where $H$ is the
pre-averaged motility matrix (an $n \times n$ non-singular matrix independent of
$t$). The noise vector $\bm{\xi}_H$ satisfies the FDR \cite[Sec.3.3]{doi86}
\begin{equation}
  \label{e.<xi-xi>.svec.HI}
  \left\langle \bm{\xi}_H(t_1) \bm{\xi}_H(t_2) \right\rangle
  =
  2k_BT \delta (t_1-t_2) I_3H.
\end{equation}
If the HI is absent, $H$ is a diagonal matrix, resulting in an elastic network
with beads of different masses \cite{yuan24}.  In particular, when
$H = \bm{I}_N / \gamma$, Eqs.~(\ref{e.rGNM.svec.HI}) and
(\ref{e.<xi-xi>.svec.HI}) reduce to Eqs.~(\ref{e.rGNM.svec}) and
(\ref{e.<xi-xi>.svec}), respectively.

It can be readily shown that Eq.~(\ref{e.(Lr)'}) also holds for $L_H$:
\begin{equation}
  \label{e.(LHr)'}
  (L_H \cdot \bm{r})'
  = L_H' \cdot (\bm{r}' - \bm{r}_i \bm{1}).
\end{equation}
Note that while $L$ is symmetric, $L_H$ is not. Therefore, $\bm{1}$ is the right
eigenvector of $L_H$ with a zero eigenvalue, but it is not necessarily the left
eigenvector. However, $L_H$ has a left eigenvector $\bm{w}$ with a zero
eigenvalue.
Because $H$ is non-singular, $\bm{w}\cdot H = \bm{1}$ if the zero mode $\bm{w}$
is suitably normalized. We assume that the $i$th entry of $\bm{w}$, $w_i$, is
not zero: $w_i \neq 0$.

\section {Derivation of three GLEs}\label{s.derivation-gle}

It is well-known that memory effects in GLEs can be characterized either by
resistance (friction) or mobility \cite{kubo91}. In this section, we derive two
GLEs: one with the resistance kernel and the other with the mobility kernel. The
interrelation of these two representations will be discussed in
Secs.~\ref{s.interrelation1} and \ref{s.interrelation2}. Additionally, using a
projection operator scheme, we also derive another GLE.

\subsection {GLE with resistance kernel}\label{s.derivation_gle_r}

By applying the prime operator to Eq.~(\ref{e.rGNM.svec}), the equations of
motion for the beads $\bm{r}_m\,(m \neq i)$ are given by
\begin{equation}
  \gamma\frac {d \bm{r}'(t)}{dt} 
  \label{e.rGNM.svec.reduced}
  =
  - L' \cdot [\bm{r}'(t) - \bm{r}_i(t) \bm{1}] +  \bm{\xi}'(t),
\end{equation}
where Eq.~(\ref{e.(Lr)'}) is used. From Eq.~(\ref{e.rGNM.1}), the equation of
motion for the $i$th bead is given by
\begin{align}
  \gamma\frac {d \bm{r}_i(t)}{dt} 
  &
  =
  \sum_{n=1}^{N} k_{in} [\bm{r}_n(t) - \bm{r}_i(t)] + \bm{\xi}_i(t)
  \notag\\[0.1cm]
  \label{e.rGNM.tagged.reduced}
  &= \bm{1}\cdot L' \cdot [\bm{r}'(t) - \bm{r}_i(t) \bm{1}] + \bm{\xi}_i(t),
\end{align}
where we used the fact that $k_{in} = -l_{in}$ if $n\neq i$ and
Eq.~(\ref{e.ell_i}). Now, the translational symmetry of the system is clearly
discernible in Eqs.~(\ref{e.rGNM.svec.reduced}) and
(\ref{e.rGNM.tagged.reduced}).

A formal solution of Eq.~(\ref{e.rGNM.svec.reduced}) is given by \vspace*{-.1cm}
\begin{align}
  \label{e.rGNM.svec.reduced.solution}
  \bm{r}'&(t)- \bm{r}_i(t) \bm{1}
  =  
  -\int_0^t e^{-\frac {L'}{\gamma}(t-\tau)} \cdot \bm{1} \dot{\bm{r}}_i(\tau) d\tau 
  \notag\\[0.1cm]
  &+\frac {1}{\gamma}\int_0^t e^{-\frac {L'}{\gamma}(t-\tau)} \cdot\bm{\xi}'(\tau)d\tau
  + e^{-\frac {L'}{\gamma}t} \cdot \delta\bm{r}'(0),
\end{align}
where integration by parts is used to obtain the first term on the right-hand
side, and $\delta\bm{r}'(0)$ is defined by
$\delta\bm{r}'(0):= \bm{r}'(0) - \bm{r}_i(0)\bm{1}$. We also utilized the fact
that the order of multiplications between vectors and symmetric matrices, such as
$e^{-L't/\gamma}$, is immaterial.

Inserting Eq.~(\ref{e.rGNM.svec.reduced.solution}) into
Eq.~(\ref{e.rGNM.tagged.reduced}), we obtain a GLE
\begin{equation}
  \label{e.gle.w.registance-kernel}
  \gamma\frac {d \bm{r}_i(t)}{dt} +\gamma\int_0^t \mu(t-\tau) \dot{\bm{r}}_i(\tau) d\tau 
  =
  \bm{\xi}_i(t) + \bm{\xi}_i^{\mathrm{r}}(t),
\end{equation}
where $\mu(t)$ is a memory kernel and $\bm{\xi}_i^r(t)$ is a colored noise
defined respectively as
\begin{align}
  \label{e.mu.1}
  \gamma\mu(t) &:= \bm{1} \cdot L' e^{-\frac {L'}{\gamma}t} \cdot \bm{1},
  \\[0.1cm]
  \label{e.xi^r.1}
  \bm{\xi}_i^{\mathrm{r}}(t) &:=
  \bm{1} \cdot L' \cdot
  \left[
  \frac {1}{\gamma} \int_0^t e^{-\frac {L'}{\gamma}(t-\tau)} \cdot \bm{\xi}'(\tau)d\tau
  +
  e^{-\frac {L'}{\gamma}t} \cdot \delta\bm{r}'(0)
  \right].
\end{align}
Thus, the interaction forces exerted on the tagged bead by the other beads are
divided into the resistance force with memory and the colored noise as
illustrated in Fig.~\ref{f.model}(b). Therefore, the memory kernel
$\gamma\mu(t)$ is referred to as a resistance kernel. We place $\gamma$ in front
of the second term in the left-hand side of
Eq.~(\ref{e.gle.w.registance-kernel}) to render the Laplace transform of
$\mu(t)$ dimensionless.  Note also that the two noise terms, the white noise
$\bm{\xi}_i(t)$ and the colored noise $\bm{\xi}_i^{\mathrm{r}}(t)$, are mutually
independent.

Equation (\ref{e.gle.w.registance-kernel}) is an overdamped version of the
GLE. This type of GLE was originally derived in Ref.~\cite{panja10b} for
single-bead motion in linear polymer models by using normal modes. Here, we
derive the same equation for a general setting without relying on normal
modes. It is also noteworthy that Eq.~(\ref{e.gle.w.registance-kernel}) with any
resistance kernel $\mu(t)$ can be numerically integrated by using a Markov
embedding method proposed in Ref.~\cite{miyaguchi22}.

Now, let us show that Eqs.~(\ref{e.mu.1}) and (\ref{e.xi^r.1}) satisfy the
FDR. From Eq.~(\ref{e.<xi-xi>.svec}), $\bm{\xi}'(t)$ satisfies the FDR
\begin{equation}
  \label{e.<xi-xi>'.svec}
  \left\langle \bm{\xi}'(t_1) \bm{\xi}'(t_2) \right\rangle
  =
  2k_BT \gamma \delta (t_1-t_2) I_3I_{N-1}.
\end{equation}
Moreover, if we assume that $\bm{r}'(t)$ is in equilibrium at $t=0$, then
$\delta\bm{r}'(0)$ follows the canonical distribution
\begin{equation}
  \label{e.canonical_r'}
  \propto
  \exp\left[
  -\frac {
    \delta\bm{r}'(0) \cdot L' I_3 \cdot \delta\bm{r}'(0)
  }{2k_BT}
  \right],
\end{equation}
where $\delta\bm{r}' \cdot L' I_3 \cdot \delta\bm{r}'/2$ is the potential energy
[See also Eq.~(\ref{e.rGNM.svec.reduced})]. In Eq.~(\ref{e.canonical_r'}), the
dot products should be understood as $3(N-1)$-dimensional contractions.
Therefore, the covariant matrix of $\delta\bm{r}'(0)$ is given by
\begin{equation}
  \label{e.<r'(0)r'(0)>}
  \left\langle \delta\bm{r}'(0) \delta\bm{r}'(0) \right\rangle
  =
  k_BT L'^{-1} I_3.
\end{equation}
Note that the inversion $L'^{-1}$ exists because $L'$ is positive definite [See
Appendix \ref{s.positiv-definite}].  From Eqs.~(\ref{e.xi^r.1}),
(\ref{e.<xi-xi>'.svec}) and (\ref{e.<r'(0)r'(0)>}), the FDR is readily obtained
as
\begin{equation}
  \label{e.<xi^r(t)xi^r(t')}
  \left\langle \bm{\xi}_i^{\mathrm{r}}(t_1) \bm{\xi}_i^{\mathrm{r}}(t_2) \right\rangle
  =
  \gamma k_BT I_3 \mu(t_1-t_2).
\end{equation}

\subsection {GLE with mobility kernel}\label{s.gle-with-m-kernel}

Next, to derive a GLE with the mobility-kernel representation \cite{sakaue13},
let us introduce an external force $\bm{f}_{\mathrm{ex}}(t)$ exerted on the
tagged bead $i$. We assume that the external force $\bm{f}_{\mathrm{ex}}(t)$ is
switched on at $t=0$ at which the system is in equilibrium. The equation of
motion is then written as
\begin{equation}
  \label{e.eom.svec.fex}
  \gamma\frac {d \bm{r}(t)}{dt} 
  =
  - L \cdot \bm{r}(t)  + \bm{f}_{\mathrm{ex}}(t) \bm{e}_i +  \bm{\xi}(t),
\end{equation}
where $\bm{e}_i$ is an $N$-dimensional unit vector
$\bm{e}_i := (0,\dots,\overset{(i)}{1},\dots,0)$, and the external force
$\bm{f}_{\mathrm{ex}}(t)$ is a three-dimensional vector.

A formal solution of Eq.~(\ref{e.eom.svec.fex}) is given by
\begin{equation}
  \bm{r}(t) =
  \frac {1}{\gamma}
  \int_0^te^{-\frac {L}{\gamma}(t-\tau)} \cdot 
  \left[
  \bm{f}_{\mathrm{ex}}(\tau) \bm{e}_i + \bm{\xi}(\tau)
  \right] 
  d\tau
  + e^{-\frac {L}{\gamma}t} \cdot \bm{r}(0). 
\end{equation}
Taking a contraction of this equation with $\bm{e}_i$ and differentiating the
result in terms of $t$, we have a GLE with the mobility-kernel representation as
\begin{equation}
  \label{e.gle.w.mobility-kernel}
  \gamma\frac {d\bm{r}_i(t)}{dt}
  =
  \bm{f}_{\mathrm{ex}}(t) +
  \int_0^t \psi(t-\tau) \bm{f}_{\mathrm{ex}}(\tau) d\tau
  + \bm{\xi}_i + \bm{\xi}_i^{\mathrm{m}}, 
\end{equation}
where $\psi(t)$ is a mobility kernel and $\bm{\xi}_i^{\mathrm{m}} (t)$ is a
correlated noise. They are defined by
\begin{align}
  \label{e.psi.1}
  \gamma\psi(t) &:= -\bm{e}_i \cdot L e^{-\frac {L}{\gamma}t} \cdot \bm{e}_i,
  \\[0.1cm]
  \label{e.xi^m.1}
  \bm{\xi}_i^{\mathrm{m}}(t) &:=
  - \bm{e}_i \cdot L
  \left[
  \frac {1}{\gamma} \int_0^t
  e^{-\frac {L}{\gamma}(t-\tau)} \cdot \bm{\xi}(\tau)d\tau
  +
  e^{-\frac {L}{\gamma}t} \cdot \delta\bm{r}(0)
  \right]
\end{align}
where we set $\delta \bm{r}(0):= \bm{r}(0) - \bm{r}_i(0) \bm{1}_N$ and use
$L\cdot \bm{1}_N = 0$ [Eq.~(\ref{e.sum_l_mn=0})] in obtaining the last term in
Eq.~(\ref{e.xi^m.1}). The equation (\ref{e.gle.w.mobility-kernel}) was derived
for the linear polymer models such as the Rouse model in Refs.~\cite{sakaue13,
  saito15} by using normal mode techniques, but here we derive it without using
normal modes for more general systems, i.e., the linear elastic network.

The resemblance of Eqs.~(\ref{e.gle.w.mobility-kernel})--(\ref{e.xi^m.1}) with
Eqs.~(\ref{e.gle.w.registance-kernel})--(\ref{e.xi^r.1}) is remarkable. For
example, the vector $\bm{1}$ and matrix $L'$ in Eq.~(\ref{e.mu.1}) correspond to
$\bm{e}_i$ and $L$ in Eq.~(\ref{e.psi.1}). Moreover, as in the case of the
resistance-kernel representation [Eq.~(\ref{e.gle.w.registance-kernel})], there
are a white noise $\bm{\xi}_i(t)$ and a correlated noise
$\bm{\xi}_i^{\mathrm{m}}(t)$ in Eq.~(\ref{e.gle.w.mobility-kernel}). However, in
this case, the two noise terms are not mutually independent, simply because
$\bm{\xi}_i^{\mathrm{m}}(t)$ in Eq.~(\ref{e.xi^m.1}) contains $\bm{\xi}_i(t)$.

To confirm the FDR, we need to derive the covariant matrix
$\left\langle \delta\bm{r}(0) \delta\bm{r}(0) \right\rangle$.
Moreover, let us recall the assumption that the beads are in equilibrium at
$t=0$, at which the external force $\bm{f}_{\mathrm{ex}}(t)$ is switched
on. Under these assumptions,
$\left\langle \delta\bm{r}'(0) \delta\bm{r}'(0) \right\rangle$ is again given by
Eq.~(\ref{e.<r'(0)r'(0)>}). Therefore, we obtain
\begin{equation}
  \label{e.<r(0)r(0)>}
  \left\langle \delta\bm{r}(0) \delta\bm{r}(0) \right\rangle
  =
  k_BT M I_3,
\end{equation}
where $M$ is an $N\times N$ matrix; $M$ is obtained by inserting zero vectors
into the $i$th row and column of $L'^{-1}$. This matrix $M$ is represented by a
block matrix representation with a non-singular matrix $J$ as
\begin{equation}
  \label{e.JMJ}
  JMJ^t
  =
  \begin{bmatrix*}[c]
    L'^{-1}  & \bm{0} \\
    \bm{0}^t & 0
  \end{bmatrix*},
\end{equation}
where $J^t$ is the transposed matrix of $J$, and $\bm{0}$ and $\bm{0}^t$ is a
column vector and a row vector, respectively. The matrix $J$ ($J^t$) moves the
$i$th row (column) of a matrix to the $N$th row (column), and $j$th row (column)
with $i+1 \leq j\leq N$ to $(j-1)$th row (column), if it is multiplied to that
matrix from the left (right). Note also that $AJ^tJB=AB$ for any $N \times N$
matrices $A$ and $B$ [See Appendix \ref{s.AJ2B=AB}].

Although $L$ does not have inversion, $M$ plays the role of inversion as follows
\cite{shinkai20}:
\begin{align}
  (JLJ^t)(JMJ^t)(JLJ^t)
  &=
  \begin{bmatrix*}[c]
    L'          & \bm{l}_i' \\
    \bm{l}_i'^t & \bm{l}_i' \cdot L'^{-1} \cdot \bm{l}_i' 
  \end{bmatrix*}
  \notag\\[0.1cm]
  \label{e.LML0}
  &=
  \begin{bmatrix*}[c]
    L'          & \bm{l}_i' \\
    \bm{l}_i'^t & l_{ii}
  \end{bmatrix*}
  =
  JLJ^t,
\end{align}
where we use $\bm{l}_i' = -L'\cdot \bm{1}$ [Eq.~(\ref{e.ell_i})] in the second
equality. It follows that
\begin{equation}
  \label{e.LML}
  LML = L.
\end{equation}
By using Eqs.~(\ref{e.xi^m.1}) and (\ref{e.LML}), it is easy to show the FDR of
the second kind
\begin{align}
  &\left\langle
  [\bm{\xi}_i(t_1)+\bm{\xi}_i^{\mathrm{m}}(t_1)]
  [\bm{\xi}_i(t_2)+\bm{\xi}_i^{\mathrm{m}}(t_2)]
  \right\rangle\notag\\[0.1cm]
  \label{e.<(xi+xi^m)(xi+xi^m)>}
  &=
  \gamma k_BTI_3[2\delta(t_1-t_2) + \psi(t_1-t_2)].
\end{align}
Note that, if $t_1 < t_2$, we have
$\left\langle \bm{\xi}_i^{\mathrm{m}}(t_1) \bm{\xi}_i(t_2) \right\rangle = 0$,
whereas
$\left\langle \bm{\xi}_i(t_1) \bm{\xi}_i^{\mathrm{m}}(t_2) \right\rangle =
2\gamma k_BTI_3\psi(t_1-t_2)$ due to the mutual dependence of the two noise
terms.

If the external force $\bm{f}_{\mathrm{ex}}(t)$ is absent, the left-hand side of
Eq.~(\ref{e.<(xi+xi^m)(xi+xi^m)>}) is equivalent to the velocity autocorrelation
$\gamma^2 \langle \bm{v}_i(t_1) \bm{v}_i(t_2) \rangle$ [See
Eq.~(\ref{e.gle.w.mobility-kernel})], where $\bm{v}_i(t)$ is the velocity of the
$i$th bead defined by $\bm{v}_i = d \bm{r}_i /dt$.  Thus, in the absence of
$\bm{f}_{\mathrm{ex}}(t)$, $\psi(t)$ is a velocity autocorrelation function at
long time apart from a constant factor as
\begin{equation}
  \label{e.<vi(t)vi(t')>}
  \left\langle \bm{v}_i(t) \bm{v}_i(0) \right\rangle
  =
  \frac {k_BT I_3}{\gamma} [2\delta (t) + \psi (t)].
\end{equation}
Therefore the negative sign in the right-hand side of Eq.~(\ref{e.psi.1})
indicates that the velocity autocorrelation is antipersistent at $t>0$, because
$L$ is positive semidefinite.  A formula for the mean square displacement (MSD)
can be readily obtained from Eq.~(\ref{e.<vi(t)vi(t')>}).

\subsection {Interrelation of two representations}\label{s.interrelation1}

Here , it is shown that the GLE with the resistance kernel
[Eq.~(\ref{e.gle.w.fex}) below] can be transformed into the GLE with the
mobility kernel [Eq.~(\ref{e.gle.w.mobility-kernel})] by using Laplace
transforms.  If there is a time-dependent external force
$\bm{f}_{\mathrm{ex}}(t)$ exerted on the tagged bead $\bm{r}_i$, the GLE with
the resistance kernel in Eq.~(\ref{e.gle.w.registance-kernel}) is given by
\begin{equation}
  \label{e.gle.w.fex}
  \gamma\frac {d \bm{r}_i(t)}{dt} +\gamma\int_0^t \mu(t-\tau) \dot{\bm{r}}_i(\tau) d\tau 
  =
  \bm{f}_{\mathrm{ex}}(t) + \bm{\xi}_i + \bm{\xi}_i^{\mathrm{r}}. 
\end{equation}
The Laplace transform of a function $\phi (t)$ is defined and denoted by
$\hat{\phi}(s) := \int_0^{\infty} e^{-ts} \phi(t) dt$.  Then, by the Laplace
transform of Eq.~(\ref{e.gle.w.fex}), we have
\begin{equation}
  \gamma\hat{\bm{v}_i}(s)
  \label{e.hat_v(s)}
  =
  \frac {\hat{\bm{f}}_{\mathrm{ex}}(s)}{1+\hat{\mu}(s)}
  +
   \frac {\hat{\bm{\xi}}_i(s) + \hat{\bm{\xi}}_i^{\mathrm{r}}(s)}{1+\hat{\mu}(s)}.
\end{equation}

To confirm the FDR for Eq.~(\ref{e.hat_v(s)}), a double Laplace transform is
defined by
\begin{equation}
  \label{e.double-laplace}
  \langle \hat{\bm{\xi}}_i(s_1)\hat{\bm{\xi}}_i(s_2) \rangle
  :=
  \int_0^{\infty}dt_1 \int_0^{\infty}dt_2 e^{-t_1s_1}e^{-t_2s_2}
  \langle \bm{\xi}_i(t_1) \bm{\xi}_i(t_2) \rangle.
\end{equation}
Inserting Eq.~(\ref{e.<xi-xi>}) into Eq.~(\ref{e.double-laplace}), we have
$\langle \hat{\bm{\xi}}_i(s_1)\hat{\bm{\xi}}_i(s_2) \rangle = 2\gamma k_BT I_3
/(s_1+s_2)$.  Similarly, the double Laplace transform of
Eq.~(\ref{e.<xi^r(t)xi^r(t')}) is given by
$\langle \hat{\bm{\xi}}_i^{\mathrm{r}}(s_1)\hat{\bm{\xi}}_i^{\mathrm{r}}(s_2)
\rangle = \gamma k_BT I_3 [\hat{\mu}(s_1) + \hat{\mu}(s_2)]/ (s_1+s_2)$
\cite{pottier03}. From these relations, we obtain the FDR for
Eq.~(\ref{e.hat_v(s)}) in the Laplace domain
\begin{align}
  &
  \left\langle
  \frac
  {\hat{\bm{\xi}}_i(s_1) + \hat{\bm{\xi}}_i^{\mathrm{r}}(s_1)}
  {1+\hat{\mu}(s_1)}
  \frac
  {\hat{\bm{\xi}}_i(s_2) + \hat{\bm{\xi}}_i^{\mathrm{r}}(s_2)}
  {1+\hat{\mu}(s_2)}
  \right\rangle\notag\\[0.1cm]
  \label{e.<[w+w^c][w+w^c]>}
  &=
  \gamma k_BTI_3
  \frac
  {\frac {1}{1+\hat{\mu}(s_1)} + \frac {1}{1+\hat{\mu}(s_2)}}
  {s_1+s_2}.
\end{align}
where we used the fact that $\bm{\xi}_i$ and $\bm{\xi}_i^{\mathrm{r}}$ are
mutually independent.

Thus, the Laplace inversion of Eq.~(\ref{e.hat_v(s)}) yields a GLE with the
mobility-kernel, which satisfies the FDR. Considering that the mobility kernel
is the velocity autocorrelation function, it should be unique. Consequently, we
conclude that Eq.~(\ref{e.hat_v(s)}) is equivalent to the Laplace transform of
Eq.~(\ref{e.gle.w.mobility-kernel})
\begin{equation}
  \gamma\hat{\bm{v}_i}(s)
  \label{e.hat_v(s).2}
  =
  [1+\hat{\psi}(s)]\hat{\bm{f}}_{\mathrm{ex}}(s)
  +
  \hat{\bm{\xi}}_i(s) + \hat{\bm{\xi}}_i^{\mathrm{m}}(s).
\end{equation}

Comparing Eqs.~(\ref{e.hat_v(s)}) and (\ref{e.hat_v(s).2}), we have a relation
between the resistance and the mobility kernels
\begin{equation}
  \label{e.pshi(t).mobility-kernel}
  1+\hat{\psi}(s) = \frac {1}{1+ \hat{\mu}(s)},
\end{equation}
and a relation between the corresponding correlated noises
\begin{equation}
  \label{e.xi_i^m.laplace}
  \hat{\bm{\xi}}_i^{\mathrm{m}}(s)
  =
  \hat{\psi}(s)\hat{\bm{\xi}}_i(s)+
  [1+\hat{\psi}(s)]\hat{\bm{\xi}}_i^{\mathrm{r}}(s).
\end{equation}
In Sec.~\ref{s.interrelation2}, in addition to the indirect proof given here,
Eq.~(\ref{e.pshi(t).mobility-kernel}) is directly derived using normal modes.

\subsection {Relation with projection operator formalism}\label{s.projection}

Here, we derive another GLE using the projection operator method.  The time
evolution operator of the Langevin equation is given by the Hermitian conjugate
of the Smoluchowski operator
$\mathcal{L}_S^{\dagger} := k_BT/\gamma (\nabla - \beta[\nabla \Phi])\cdot
\nabla$ \cite[Eq.~(6.38)]{dhont96}, where $\Phi$ is the potential energy and
$\beta = 1/k_BT$. Because the potential $\Phi$ is arbitrary, the projection
operator method can be applied to systems with nonlinear interactions such as
those involving excluded volume effects. It is also possible to incorporate the
HI into the operator $\mathcal{L}_S^{\dagger}$; however, it is not considered
here.

The operator $\mathcal{L}_S^{\dagger}$ for the linear elastic network without
the HI in Eq.~(\ref{e.rGNM.svec}) is given by
\begin{equation}
  \label{e.L^dagger}
  \mathcal{L}_S^{\dagger}
  =
  \frac {k_BT}{\gamma} \nabla^2 -
  \frac {1}{\gamma} (LI_3): \bm{r} \nabla,
\end{equation}
where $\nabla$ is the $3N$-dimensional gradient operator with respect to
$\bm{r}$, and the double contractions, indicated by '$:$', are taken over the
$3N$-element pairs including $I_3$ (i.e., $\bm{r}$ should be regarded as a
$3N$-dimensional vector).

Furthermore, we assume that the elastic network is confined by a spherically
symmetric potential $\mathcal{U}(\bm{r}) = \sum_{k=1}^N U(r_k)$ to ensure the
existence of an equilibrium state, which is necessary to define an inner product
below.  In particular, we employ a square-well potential such that $U(r)=0$ for
$r<R$, and $U(r)=\infty$ for $r>R$ with sufficiently large $R$. As a result of
the spherical symmetry, the equilibrium state is statistically isotropic.

Now, let us define a projection operator $\mathcal{P}$ as
\cite[Eq.~(9.1.1)]{hansen90}
\begin{equation}
  \label{e.projection_P}
  \mathcal{P} \bm{A}(\bm{r})
  :=
  (\bm{r}_i, \bm{A}(\bm{r})) (\bm{r}_i, \bm{r}_i)^{-1} \bm{r}_i,  
\end{equation}
where $\bm{A}(\bm{r})$ is a three-dimensional vector function of the phase space
coordinates $\bm{r}$, the inner product $(\bullet, \bullet)$ is a $3\times 3$
tensor defined by
\begin{equation}
  \label{e.inner_product}
  (\bm{A}, \bm{B})
  :=
  \int \bm{A}^{\ast} (\bm{r}) \bm{B}(\bm{r}) P_{\mathrm{eq}}(\bm{r}) d\bm{r},
\end{equation}
$(\bm{r}_i, \bm{r}_i)^{-1}$ is the inversion of $(\bm{r}_i, \bm{r}_i)$, and
$P_{\mathrm{eq}}(\bm{r})$ is an equilibrium distribution.  In
Eq.~(\ref{e.inner_product}), $\bm{A}^{\ast}$ is the complex conjugate of
$\bm{A}$. Moreover, the projection to the orthogonal complementary space is
denoted as $\mathcal{Q}$, which is defined by
\begin{equation}
  \label{e.projection_Q}
  \mathcal{Q} := \mathcal{I} - \mathcal{P},
\end{equation}
with $\mathcal{I}$ being an identity operator.

Let us define a function $\bm{c}(t, \bm{r})$ as
\begin{equation}
  \label{e.c(t)}
  \bm{c}(t, \bm{r}) := e^{\mathcal{L}_S^{\dagger}t} \bm{r}_i.
\end{equation}
Note that $\bm{c}(t, \bm{r})$ is closely related to $\bm{r}_i(t)$, but they are
not equivalent, because the latter is a random process while the former is
not. Thus, we employ the different notation $\bm{c}(t,\bm{r})$ instead of
$\bm{r}_i(t)$. Hereafter, we denote it as $\bm{c}(t)$ instead of
$\bm{c}(t,\bm{r})$ for brevity. As is well known, $\bm{c}(\bm{r})$ follows a
GLE \cite[Eq.~(6.167)]{dhont96}
\begin{equation}
  \label{e.dc/dt}
  \gamma\frac {\partial \bm{c}}{\partial t}
  =
  \gamma e^{\mathcal{L}_S^{\dagger}t}\mathcal{P}\mathcal{L}_S^{\dagger} \bm{r}_i 
  +
  \gamma\int_0^t \bm{M}(t-\tau)\cdot\bm{c}(\tau)d\tau
  +
  \bm{\xi}^{\mathrm{pr}}(\bm{r},t),
\end{equation}
where the memory kernel $\bm{M}(t)$ and the fluctuating force
$\bm{\xi}^{\mathrm{pr}}(\bm{r},t)$ are defined by
\begin{align}
  \label{e.M(t).def}
  \bm{M}(t)
  &:= \frac 1{\gamma^2} (\bm{\xi}^{\mathrm{pr}}(t), \bm{\xi}^{\mathrm{pr}}(0))
  (\bm{r}_i, \bm{r}_i)^{-1},\\[0.1cm]
  \label{e.xi^pr.def}
  \bm{\xi}^{\mathrm{pr}}(t)
  &:=
  \gamma
  e^{\mathcal{Q} \mathcal{L}_S^{\dagger}t}
  \mathcal{Q} \mathcal{L}_S^{\dagger} \bm{r}_i. 
\end{align}
These equations (\ref{e.M(t).def}) and (\ref{e.xi^pr.def}) represents the FDR
for the GLE in Eq.~(\ref{e.dc/dt}).

It is readily checked that
\begin{equation}
  \mathcal{L}_S^{\dagger} \bm{r}_j = - \frac {1}{\gamma} \bm{l}_j \cdot \bm{r},
\end{equation}
where the contraction on the right-hand side is taken over the $N$-element pairs
(i.e., $\bm{r}$ is considered as an $N$-dimensional supervector. In the
following, the dot product should be treated in the same way).  It follows that
\begin{equation}
  \label{e.PLr_j}
  \mathcal{P} \mathcal{L}_S^{\dagger} \bm{r}_j
  =
  -\frac {l_{ij}}{\gamma} \bm{r}_i,
\end{equation}
where we used the fact that the system is isotropic, and therefore
$(\bm{r}_k, \bm{r}_i)=0$ if $k\neq i$. By setting $j=i$ in Eq.~(\ref{e.PLr_j})
and applying $e^{\mathcal{L}_S^{\dagger}t}$ from the left, we have
\begin{equation}
  \label{e.PLr_i}
  e^{\mathcal{L}_S^{\dagger}t} \mathcal{P} \mathcal{L}_S^{\dagger} \bm{r}_i
  =
  -\frac {\bm{1}\cdot L' \cdot \bm{1}}{\gamma} \bm{c}(t)
  =
  -\mu(0)\bm{c}(t)
\end{equation}
where we used Eqs.~(\ref{e.ell_i}) and (\ref{e.mu.1}).  The equation
(\ref{e.PLr_i}) is an explicit expression of the first term on the right-hand
side of Eq.~(\ref{e.dc/dt}).

By using Eqs.~(\ref{e.projection_Q}) and (\ref{e.PLr_j}), we also have a
projection on the orthogonal complementary space as
\begin{equation}
  \label{e.QLr_j}
  \mathcal{Q} \mathcal{L}_S^{\dagger} \bm{r}_j
  =
  \mathcal{L}_S^{\dagger} \bm{r}_j
  -
  \mathcal{P} \mathcal{L}_S^{\dagger} \bm{r}_j
  =
  -\frac {1}{\gamma} \bm{l}'_j\cdot\bm{r}'.
\end{equation}
In particular, for $j=i$, the Eq.~(\ref{e.QLr_j}) is rewritten as
\begin{equation}
  \label{e.QLr_i}
  \mathcal{Q} \mathcal{L}_S^{\dagger} \bm{r}_i
  =
  \frac {1}{\gamma} \bm{1} \cdot L' \cdot \bm{r}',
\end{equation}
where Eq.~(\ref{e.ell_i}) is used.  Moreover, combining Eq.~(\ref{e.QLr_j}) for
all $j \neq i$, we obtain an important expression
\begin{equation}
  \label{e.QLr'}
  \mathcal{Q} \mathcal{L}_S^{\dagger} \bm{r}'
  =
  -\frac {1}{\gamma} L'\cdot\bm{r}'.
\end{equation}
These relations (\ref{e.QLr_i}) and (\ref{e.QLr'}) clearly show that $L'$ is
closely related to $\mathcal{Q} \mathcal{L}_S^{\dagger}$, which is a
time-evolution operator in the orthogonal complementary space.

By using Eqs.~(\ref{e.QLr_j}) and (\ref{e.QLr'}), we obtain an explicit
expression for the fluctuating force as
\begin{equation}
  \label{e.xi^pr}
  \bm{\xi}^{\mathrm{pr}}(t)
  =
  \bm{1}\cdot L' e^{-\frac {L'}{\gamma}t} \cdot \bm{r}'.
\end{equation}
Unexpectedly, Eq.(\ref{e.xi^pr}) is not equivalent to the non-thermal part of
Eq.~(\ref{e.xi^r.1}); this inconsistency shall be resolved shortly. Then,
inserting Eq.~(\ref{e.xi^pr}) into the definition of $\bm{M}(t)$ in
Eq.~(\ref{e.M(t).def}), and using a relation
$(\bm{r}', \bm{r}')= (\bm{r}_i, \bm{r}_i)I_{N-1}$ due to the isotropy, the
memory kernel $\bm{M}(t)$ is given by
\begin{equation}
  \label{e.M(t)}
  \bm{M}(t) = - \frac {d\mu(t)}{dt} I_3,
\end{equation}
where $\mu(t)$ is the resistance kernel given by Eq.~(\ref{e.mu.1}).

Inserting Eqs.~(\ref{e.PLr_i}), (\ref{e.xi^pr}) and (\ref{e.M(t)}) into
Eq.~(\ref{e.dc/dt}), and then performing integration by parts, we have an
equation for $\bm{c}(t)$ explicitly as
\begin{equation}
  \label{e.dc/dt.explicit}
  \gamma\frac {\partial \bm{c}(t)}{\partial t}
  +
  \gamma\int_0^t \mu(t-\tau)\dot{\bm{c}}(\tau)d\tau
  =
  \bm{\xi}^{\mathrm{pr}}(t)
  -\gamma\mu(t) \bm{c}(0).
\end{equation}

By using Eqs.~(\ref{e.mu.1}) and (\ref{e.xi^pr}), the right-hand side of
Eq.~(\ref{e.dc/dt.explicit}) is rewritten as
\begin{equation}
  \label{e.dc/dt.noise}
  \bm{1}\cdot L' e^{-\frac {L'}{\gamma}t} \cdot
  [\bm{r}' - \bm{c}(0)\bm{1} ],
\end{equation}
which is equivalent to the non-thermal part of Eq.~(\ref{e.xi^r.1}), because
$\bm{c}(0) = \bm{r}_i$. Thus, although Eq.~(\ref{e.dc/dt.explicit}) is similar
to the GLE with the resistance kernel in Eq.~(\ref{e.gle.w.registance-kernel}),
the thermal noise terms $\bm{\xi}_i$ and $\bm{\xi}'$ are absent as
expected. Moreover, the term with $\bm{c}(0)\bm{1}$ in Eq.~(\ref{e.dc/dt.noise})
is not a vector in the orthogonal complementary space, and thus it is not
included in the fluctuating force $\bm{\xi}^{\mathrm{pr}}$ in the projection
operator scheme. In contrast, the corresponding term $\bm{r}_i(0) \bm{1}$ in
Eq.~(\ref{e.gle.w.registance-kernel}) is included in the noise term
$\bm{\xi}^{\mathrm{r}}_i$ [Eq.~(\ref{e.xi^r.1})].

To obtain an example of an evolution equation, let us multiply $\bm{r}_i/\gamma$
with Eq.~(\ref{e.dc/dt.explicit}) and take an ensemble average in terms of
$P_{\mathrm{eq}}(\bm{r})$, thereby obtaining \cite[Eq.~(6.150)]{dhont96}
\begin{align}
  \frac {d \left\langle \bm{r}_i(t) \bm{r}_i(0) \right\rangle}{dt}
  &+
  \int_0^t \mu(t-\tau) \left\langle \dot{\bm{r}}_i(\tau) \bm{r}_i(0) \right\rangle d\tau.
  \notag\\[0.1cm]
  \label{e.<r(t)r(0)>}
  &=
  -\mu(t) \left\langle \bm{r}_i(0) \bm{r}_i(0) \right\rangle.
\end{align}
The same equation can be readily derived from
Eq.~(\ref{e.gle.w.registance-kernel}).

\section {Normal modes}\label{s.normal-modes}

To derive explicit expressions of the GLEs, it is necessary to utilize the
normal modes. Diagonalization of $L$ is straightforward \cite{shinkai20} (See
Sec.~\ref{s.normal-mode.mobility}), whereas that of $L'$ is somewhat intricate
to ensure translational symmetry in normal-mode equations.  In addition,
Eq.~(\ref{e.pshi(t).mobility-kernel}) is directly proved by using the normal
modes.

\subsection {Diagonalization of $L'$}\label{s.diag-L-prime}

Because $L'$ is symmetric and positive definite (See Appendix
\ref{s.positiv-definite}), there exist positive eigenvalues
$\lambda_1, \dots, \lambda_{N-1}$ and corresponding eigenvectors
$\bm{v}_1, \dots, \bm{v}_{N-1}$. That is,
$L' \cdot [\bm{v}_1, \dots, \bm{v}_{N-1}] = [\lambda_1\bm{v}_1, \dots,
\lambda_{N-1}\bm{v}_{N-1}]$. Alternatively, this relation can be expressed as
\begin{equation}
  \label{e.L'Q=QLambda}
L' Q = Q \Lambda,
\end{equation}
where an $(N-1)\times(N-1)$ non-singular matrix $Q$ is defined by
\begin{equation}
  \label{e.Q.def}
  Q := [\bm{v}_1, \dots, \bm{v}_{N-1}],
\end{equation}
and a diagonal matrix $\Lambda$ is defined by
$(\Lambda)_{jk} = \lambda_j \delta_{jk}$.

Moreover, we define constants $c_k>0$ as $c_k:= \bm{v}_k^2$, i.e., $\bm{v}_k$ is
not necessarily a unit vector as typically assumed. Instead, we impose a
condition
\begin{equation}
  \label{e.1Q=1}
\bm{1}\cdot Q = \bm{1}.
\end{equation}
This condition implies that the sum of the elements of $\bm{v}_k$ equals unity,
thereby specifying the eigenvectors $\bm{v}_k$ and the values of the constants
$c_k$.  Thanks to the condition in Eq.~(\ref{e.1Q=1}), normal-mode equations of
motion remain translationally symmetric [See Eqs.~(\ref{e.dy/dt}) and
(\ref{e.dri/dt.normal-mode})].

If the sum of the elements of $\bm{v}_k$ vanishes, we consider $c_k$ as
$\infty$. This is justified as follows. Such a mode $k$ does not contribute to
the dynamics of the tagged bead $\bm{r}_i$, due to $\bm{1}\cdot\bm{v}_k = 0$ and
Eq.~(\ref{e.mu.1}) with the spectral decomposition
$L' = \sum_j \lambda_j \bm{v}_j\bm{v}_j$. We can remove the contribution from
the mode $k$ by formally setting $c_k=\infty$ in Eq.~(\ref{e.mu(t)}) below. This
situation arises when the $k$th mode is independent of the tagged bead
$\bm{r}_i$. An example of this is the ring polymer investigated in
Sec.~\ref{s.ring-polymer}.

Left multiplication of Eq.~(\ref{e.L'Q=QLambda}) by $Q^t$ results in the
diagonalization of $L'$ as
\begin{equation}
  \label{e.Q^tL'Q=CLambda}
  Q^t L' Q = C \Lambda 
\end{equation}
where $Q^t$ is the transposed matrix of $Q$ and a diagonal matrix $C$ is defined
as $(C)_{jk} = c_j \delta_{jk}$. Because $Q^t Q = C$ and $C$ is non-singular, we
have
\begin{align}
  \label{e.Q^tQC^-1}
  Q^tQC^{-1} &= I_{N-1},\\[0.1cm]
  \label{e.QC^-1Q^t}
  QC^{-1}Q^t &= I_{N-1}.
\end{align}
Moreover, by using Eqs.~(\ref{e.Q^tL'Q=CLambda})--(\ref{e.QC^-1Q^t}), we have
\begin{equation}
  \label{e.L'=QC^-1LambdaQ^t}
  L' = QC^{-1}\Lambda Q^t. 
\end{equation}
This equation is rewritten as $Q^{-1}L' (Q^t)^{-1} = C^{-1}\Lambda$, and
therefore we obtain
\begin{equation}
  \label{e.Q^tL'^-1Q}
  Q^t L'^{-1} Q = C \Lambda^{-1}.
\end{equation}


\subsection {Normal-mode GLE with resistance kernel}

Multiplying $C^{-1}Q^t$ from the left of Eq.~(\ref{e.rGNM.svec.reduced}),
inserting the identity $QC^{-1}Q^t$ [Eq.~(\ref{e.QC^-1Q^t})] behind $L'$, and
using Eqs.~(\ref{e.1Q=1}) and (\ref{e.Q^tL'Q=CLambda}), we have
\begin{equation}
  \label{e.dy/dt}
  \gamma C^{-1} \cdot \frac {d \bm{y}(t)}{dt}
  =
  - C^{-1}\Lambda \cdot [\bm{y}(t)- \bm{r}_i(t) \bm{1}] + \bm{\eta}(t),
\end{equation}
where we define $\bm{y}$ and $\bm{\eta}$ respectively as
$\bm{y}:= Q^t\cdot\bm{r}'$ and $\bm{\eta} := C^{-1}Q^t\cdot
\bm{\xi}'$. Similarly, inserting Eq.~(\ref{e.QC^-1Q^t}) on both sides of $L'$ in
Eq.~(\ref{e.rGNM.tagged.reduced}) and using Eqs.~(\ref{e.1Q=1}) and
(\ref{e.Q^tL'Q=CLambda}), we have
\begin{equation}
  \label{e.dri/dt.normal-mode}
  \gamma \frac {d \bm{r}_i(t)}{dt}
  =
  \bm{1}\cdot C^{-1}\Lambda \cdot [\bm{y}(t)- \bm{r}_i(t) \bm{1}] + \bm{\xi}_i(t).
\end{equation}
Equations (\ref{e.dy/dt}) and (\ref{e.dri/dt.normal-mode}) form the normal-mode
equations of motion.

Importantly, the noise $\bm{\eta}(t)$ in Eq.~(\ref{e.dy/dt}) satisfies the FDR
in the form
\begin{equation}
  \label{e.<eta(t)eta(t')>}
  \left\langle \bm{\eta}(t_1) \bm{\eta}(t_2)\right\rangle
  =
  2\gamma C^{-1}k_BT I_3  \delta(t_1-t_2), 
\end{equation}
which follows from Eq.~(\ref{e.<xi-xi>'.svec}). By multiplying $Q^t$ and $Q$ to
Eq.~(\ref{e.<r'(0)r'(0)>}) respectively from the left and the right, and using
Eq.~(\ref{e.Q^tL'^-1Q}), we obtain a covariant matrix
\begin{equation}
  \label{e.<(y-r_i1)(y-r_i1)>}
  \left\langle
  \delta\bm{y}(0) \delta\bm{y}(0)
  \right\rangle
  =
  k_BT C\Lambda^{-1} I_3.
\end{equation}
where we define $\delta\bm{y}(0) := \bm{y}(0) - \bm{r}_i(0) \bm{1}$.

As in Sec.~\ref{s.derivation_gle_r}, it is possible to derive a normal-mode GLE
by formally solving Eq.~(\ref{e.dy/dt}) and inserting the result into
Eq.~(\ref{e.dri/dt.normal-mode}). But, it is straightforward to simply rewrite
Eqs.~(\ref{e.mu.1}) and (\ref{e.xi^r.1}) with the diagonal matrices $\Lambda$
and $C$.  By using Eqs.~(\ref{e.L'=QC^-1LambdaQ^t}) and (\ref{e.1Q=1}) in
Eq.~(\ref{e.mu.1}), the memory kernel $\mu(t)$ is rewritten as
\begin{equation}
  \label{e.mu(t)}
  \gamma\mu(t)
  =
  \mathrm{tr} \left(C^{-1}\Lambda e^{-\frac {\Lambda}{\gamma}t}\right)
  =
  \sum_{j=1}^{N-1} \frac {\lambda_j}{c_j} e^{-\frac {\lambda_j}{\gamma} t},
\end{equation}
where $\mathrm{tr}$ denotes the trace of a matrix. Similarly, the correlated
noise $\bm{\xi}_i^{\mathrm{r}}(t)$ in Eq.~(\ref{e.xi^r.1}) is rewritten as
\begin{equation}
  \label{e.xi_i^r}
  \bm{\xi}_i^{\mathrm{r}}(t)=
  \bm{1}\cdot C^{-1}\Lambda
  \left[
  \frac{C}{\gamma}\int_0^t e^{-\frac {\Lambda}{\gamma}(t-\tau)}\cdot\bm{\eta}(\tau)d\tau 
  +
  e^{-\frac {\Lambda}{\gamma}t} \cdot \delta\bm{y}(0)
  \right].
\end{equation}
In addition, the FDR [Eq.~(\ref{e.<xi-xi>'.svec})] for these normal-mode
representations [Eqs.~(\ref{e.mu(t)}) and (\ref{e.xi_i^r})] can be confirmed by
using Eqs.~(\ref{e.<eta(t)eta(t')>}) and (\ref{e.<(y-r_i1)(y-r_i1)>}).

Thus, we have the normal-mode GLE with the resistance kernel:
Eq.~(\ref{e.gle.w.registance-kernel}) along with Eqs.~(\ref{e.mu(t)}) and
(\ref{e.xi_i^r}). Since we have derived the GLE without utilizing the normal
mode in the previous section, the diagonalization process becomes a
straightforward problem in linear algebra.  Note also that the matrix $C$
appears in these equations instead of $Q$; this should be an advantage of the
diagonalization procedure presented in this subsection.

\subsection {Normal-mode GLE with mobility kernel}\label{s.normal-mode.mobility}

Because $L$ is symmetric, it can be diagonalized with an orthogonal matrix $P$
as $P^t L P = \Sigma$, where $\Sigma$ is a diagonal matrix defined by using
eigenvalues $\sigma_j$ as $(\Sigma)_{jk} = \sigma_j\delta_{jk}$. We set that
$\sigma_1=0$ and $\sigma_j > 0\,(j\geq2)$.  Multiplying $P^t$ to
Eq.~(\ref{e.eom.svec.fex}) from the left, we have
\begin{equation}
  \label{e.dz/dt}
  \gamma \frac {d \bm{z}(t)}{dt}
  = - \Sigma \cdot \bm{z}(t) + \bm{f}_{\mathrm{ex}}(t) P^t\cdot \bm{e}_i + \bm{\zeta}(t),
\end{equation}
where we define $\bm{z}$ and $\bm{\zeta}$ as $\bm{z}:= P^t\cdot \bm{r}$ and
$\bm{\zeta} := P^t \cdot \bm{\xi}$. The first mode $\bm{z}_1(t)$, which
corresponds to the zero eigenvalue $\sigma_1=0$ with the associated eigenvector
$\bm{1}_N/\sqrt{N}$, describes the center-of-mass motion.

Moreover, $\bm{\zeta}(t)$ in Eq.~(\ref{e.dz/dt}) satisfies the FDR
\begin{equation}
  \label{e.<zeta(t)zeta(t')>}
  \left\langle \bm{\zeta}(t) \bm{\zeta}(t') \right\rangle
  =
  2 \gamma k_BT\delta(t-t') I_3 I_N,
\end{equation}
which follows from Eq.~(\ref{e.<xi-xi>.svec}) and the orthogonality
$P^t P = I_N$. By multiplying $P^t$ and $P$ to Eq.~(\ref{e.<r(0)r(0)>})
respectively from the left and the right, and using Eq.~(\ref{e.Q^tL'^-1Q}), we
obtain a covariant matrix
\begin{equation}
  \label{e.<dz(0)dz(0)>}
  \left\langle
  \delta\bm{z}(0) \delta\bm{z}(0)
  \right\rangle
  =
  k_BT P^tMP I_3,
\end{equation}
where $\delta \bm{z}(0)$ is defined by
$\delta \bm{z}(0) := \bm{z}(0) - \bm{r}_i(0)P^t\cdot \bm{1}_N$.

It is possible to derive a normal-mode GLE with the mobility kernel by applying
$\bm{e}_i\cdot P$ to a formal solution of Eq.~(\ref{e.dz/dt}) and then
differentiating the resulting equation in terms of $t$. However, here the GLE is
obtained by simply replacing $L$ in Eqs.~(\ref{e.psi.1}) and (\ref{e.xi^m.1})
with $P\Sigma P^t$. Thus, we have
\begin{align}
  \label{e.psi.normal-mode}
  \gamma\psi(t) &=
  -\bm{e}_i \cdot P\Sigma e^{-\frac {\Sigma}{\gamma}t} P^t\cdot \bm{e}_i
  = -\sum_{j=1}^{N} p_{ij}^2\sigma_j e^{-\frac {\sigma_j}{\gamma}t},  \\[0.1cm]
  \label{e.xi^m.normal-mode}
  \bm{\xi}_i^{\mathrm{m}}(t) &=
  - \bm{e}_i \cdot P\Sigma 
  \left[
  \frac {1}{\gamma} \int_0^t
  e^{-\frac {\Sigma}{\gamma}(t-\tau)} \cdot\bm{\zeta}(\tau)d\tau
  +
  e^{-\frac {\Sigma}{\gamma}t} \cdot\delta\bm{z}(0)
  \right],
\end{align}
where $p_{ij}$ is the $(i,j)$ entry of the matrix $P$. In addition, the FDR
[Eq.~(\ref{e.<(xi+xi^m)(xi+xi^m)>})] for these normal-mode representations
[Eqs.~(\ref{e.psi.normal-mode}) and (\ref{e.xi^m.normal-mode})] can be confirmed
by using Eqs.~(\ref{e.<zeta(t)zeta(t')>}) and (\ref{e.<dz(0)dz(0)>}).  Thus, we
have the normal-mode GLE with the mobility kernel:
Eq.~(\ref{e.gle.w.mobility-kernel}) along with Eqs.~(\ref{e.psi.normal-mode})
and (\ref{e.xi^m.normal-mode}).

\subsection {Direct proof of Eq.~(\ref{e.pshi(t).mobility-kernel})}\label{s.interrelation2}

The two memory kernels in Eqs.~(\ref{e.mu(t)}) and (\ref{e.psi.normal-mode}) are
interrelated through Eq.~(\ref{e.pshi(t).mobility-kernel}). However, the
relation is not immediately obvious even from these normal-mode expressions
[Eqs.~(\ref{e.mu(t)}) and (\ref{e.psi.normal-mode})].  Here, we directly prove
Eq.~(\ref{e.pshi(t).mobility-kernel}) by utilizing a formula concerning Schur
complements.

First, Laplace transform of Eq.~(\ref{e.mu(t)}) is given by
\begin{equation}
  \label{e.mu(s).normal-mode}
  1+\hat{\mu}(s) =
  1+\sum_{j=1}^{N-1} \frac {\lambda_j}{c_j}
  \frac {1}{\tilde{s} + \lambda_{j}},
\end{equation}
where we set $\tilde{s}:= \gamma s$.  Similarly, Laplace transform of
Eq.~(\ref{e.psi.normal-mode}) is given by
\begin{equation}
  \label{e.psi(s).normal-mode}
  1+\hat{\psi}(s) =
  1-\sum_{j=1}^{N} p_{ij}^2 
  \frac {\sigma_j}{\tilde{s} + \sigma_{j}}
  =
  \sum_{j=1}^{N} p_{ij}^2
  \frac {\tilde{s}}{\tilde{s} + \sigma_{j}},
\end{equation}
where $\sum_{j=1}^{N} p_{ij}^2= 1$ is used. We show that the left-hand sides of
Eqs.~(\ref{e.mu(s).normal-mode}) and (\ref{e.psi(s).normal-mode}) are in the
reciprocal relation indicated in Eq.~(\ref{e.pshi(t).mobility-kernel}).

Let us set $A:= -L- \tilde{s} I_N$ and $A':= -L'- \tilde{s} I_{N-1}$. Using a
formula for the determinant of the Schur complement
\cite[Sec.~0.8.5.1]{horn12book}, we have
\begin{align}
  \mathrm{det} A
  &=
  \mathrm{det} A'
  \left(
  - l_{ii} - \tilde{s} - \bm{l}_i' \cdot A'^{-1} \cdot \bm{l}_i'
  \right)
  \notag\\[0.1cm]
  \label{e.schur}
  &=
  \mathrm{det} A'
  \left(
  - l_{ii} - \tilde{s} - \bm{1} \cdot L' A'^{-1} L' \cdot \bm{1}
  \right),
\end{align}
where Eq.~(\ref{e.ell_i}) is used.  Here, $A'^{-1}$ is the resolvent of $-L'$
and is calculated by $A'^{-1}=-\int_0^{\infty} e^{- \tilde{s}t} e^{-L't} dt$
\cite{dhont96}. Consequently, we have a diagonalized expression of
$L'A'^{-1}L'$ as follows:
\begin{align}
  \bm{1}\cdot L'A'^{-1}L' \cdot \bm{1}
  &= \bm{1} \cdot C^{-1}\Lambda^2 (-\Lambda - \tilde{s} I_{N-1})^{-1} \cdot \bm{1}
  \notag\\[0.1cm]
  \label{e.A'^-1}
  &= - \sum_{j=1}^{N-1} \frac {\lambda_j^2}{c_j} \frac {1}{\lambda_j + \tilde{s}},
\end{align}
where Eq.~(\ref{e.L'=QC^-1LambdaQ^t}) is utilized.

Moreover, From Eqs.~(\ref{e.sum_l_mn=0}) and (\ref{e.ell_i}), we have
$l_{ii} = \bm{1}\cdot L' \cdot \bm{1}$. Consequently, by applying the vector
$\bm{1}$ from both sides of Eq.~(\ref{e.L'=QC^-1LambdaQ^t}), we have
\begin{equation}
  \label{e.lii}
  l_{ii}
  =
  \bm{1}\cdot L' \cdot \bm{1}
  = \sum_{j=1}^{N-1} \frac {\lambda_j}{c_j}.
\end{equation}
Inserting Eqs.~(\ref{e.A'^-1}) and (\ref{e.lii}) into Eq.~(\ref{e.schur}), we
have
\begin{equation}
  \label{e.1+mu(s)}
  \frac 1{1 + \hat{\mu}(s)}
  =
  - \tilde{s} \frac {\mathrm{det}A'}{\mathrm{det}A}
  =
  - \tilde{s}(A^{-1})_{ii},
\end{equation}
where $(A^{-1})_{ii}$ is the $(i, i)$ entry of $A^{-1}$, and a formula of the
inverse matrix is used in the second equality.

Similarly, $A^{-1}$, which is the resolvent of $-L$, can be calculated as
\begin{equation}
  A^{-1} = P (- \Sigma - \tilde{s} I_N)^{-1} P^t.
\end{equation}
It follows that 
\begin{equation}
  \label{e.A^-1_ii}
  -\tilde{s}(A^{-1})_{ii}
  = \sum_{j=1}^N p_{ij}^2 \frac {\tilde{s}}{\tilde{s} + \sigma_j}
  = 1 + \hat{\psi}(s),
\end{equation}
where Eq.~(\ref{e.psi(s).normal-mode}) is used.  From Eqs.~(\ref{e.1+mu(s)}) and
(\ref{e.A^-1_ii}), we obtain Eq.~(\ref{e.pshi(t).mobility-kernel}), thereby
completing the proof.


\subsection {Asymptotics}

Here, we briefly discuss short-time and long-time diffusion of the tagged
bead. For small and large $\tilde{s}$, Eq.~(\ref{e.psi(s).normal-mode}) is
approximately expressed as
\begin{align}
  \label{e.psi(s).asympt}
  1 + \hat{\psi(s)} \simeq
  \begin{cases}
    1                   & (\tilde{s} \gg \sigma_N), \\[0.1cm]
    p_{i1}^2 = \frac 1N & (\tilde{s} \ll \sigma_2),
  \end{cases}
\end{align}
where we used $\sigma_1=0$, and $p_{i1}=1/\sqrt{N}$. $\sigma_2$ and $\sigma_N$
are the smallest and largest positive eigenvalues of $L$, respectively. Laplace
inversion of Eq.~(\ref{e.psi(s).asympt}) yields the velocity autocorrelation
[Eq.~(\ref{e.<vi(t)vi(t')>})] as
\begin{align}
  \left\langle \bm{v}_i(t) \bm{v}_i(0) \right\rangle
  \simeq
  \begin{cases}
    \frac {2k_BT}{\gamma}\delta(t)I_3   & (t \ll \gamma/\sigma_N), \\[0.1cm]
    \frac {2k_BT}{\gamma N}\delta(t)I_3 & (t \gg \gamma/\sigma_2).
  \end{cases}
\end{align}

The short-time and long-time diffusion coefficient can be obtained using the
Green-Kubo formula \cite{evans07}:
\begin{equation}
  \label{s.diffusivity_at_short_long}
  D =
  \frac {1}{3}
  \int_0^{\infty} \left\langle \bm{v}_i(t)\cdot\bm{v}_i(0) \right\rangle
  =
  \begin{cases}
    \frac {k_BT}{\gamma}   & (t \ll \gamma/\sigma_N), \\[0.1cm]
    \frac {k_BT}{\gamma N} & (t \gg \gamma/\sigma_2).
  \end{cases}
\end{equation}
The long-time diffusivity $k_BT/\gamma N$ is the diffusion coefficient for the
center of mass motion. Behaviors at intermediate timescales depend on the
details of the system.

\section {Simple polymer models}\label{s.rouse-model}

In this section, we investigate the Rouse model \cite{rouse53, doi86} and the
ring polymer as simple examples of the linear elastic network. In particular,
for the Rouse model, the memory kernels and the correlated noises depend on the
position of the tagged bead, resulting in different GLEs for the end bead and
the middle bead.

\subsection {Rouse model}\label{s.rouse-model2}
\subsubsection {Resistance kernel}

First, let us begin with a derivation of the resistance kernel $\mu(t)$.  The
interaction matrix $L$ in Eq.~(\ref{e.rGNM.svec}) for the Rouse model is given
by
\begin{equation}
  \label{e.rouse.L}
  L
  =
  \kappa
  \begin{pmatrix*}[r]
    1  & -1 & 0  & 0  &        &              \\
    -1 & 2  & -1 & 0  &        &              \\
    0  & -1 & 2  & -1 &        &              \\
       &    &    &    & \ddots &              \\
       &    &    &    &        & -1 & 2  & -1 \\
       &    &    &    &        & 0  & -1 & 1
  \end{pmatrix*},
\end{equation}
where $\kappa$ is the spring constant.  The matrix $L'$ is then obtained by
removing the $i$th row and column from $L$ as
\begin{equation}
  \label{e.rouse.L'}
  L'
  =
  \kappa
  \begin{pmatrix*}[r]
    L_1           & \text{\large 0} \\[.2cm]
   \text{\large 0} & L_2
  \end{pmatrix*},
\end{equation}
where $L_1$ is an $(i-1)\times(i-1)$ matrix and $L_2$ is an $(N-i) \times (N-i)$
matrix. They are defined respectively by
\begin{equation}
  \label{e.rouse.L1'}
  L_1
  =
  \kappa
  \begin{pmatrix*}[r]
    1  & -1 & 0  &         &              \\
    -1 & 2  & -1 &         &              \\
       &    &    &  \ddots &              \\
       &    &    &         & -1 & 2  & -1 \\
       &    &    &         & 0  & -1 & 2
  \end{pmatrix*}.
\end{equation}
and
\begin{equation}
  \label{e.rouse.L2'}
  L_2
  =
  \kappa
  \begin{pmatrix*}[r]
    2  & -1 & 0  &         &              \\
    -1 & 2  & -1 &         &              \\
       &    &    &  \ddots &              \\
       &    &    &         & -1 & 2  & -1 \\
       &    &    &         & 0  & -1 & 1
  \end{pmatrix*}.
\end{equation}

The eigenvalues and eigenvectors of $L_1$ and $L_2$ can be obtained by solving
recursion relations (See Ref.~\cite[Sec.XVI.3]{feller68} for a procedure). The
eigenvalues of $L_1$ is given by \cite{vandebroek17}
\begin{equation}
  \label{e.lambda_k}
  \lambda_k = 4\kappa \sin^2 \frac {\beta_k\pi}{2},
  \quad (k=1,\dots,i-1),
\end{equation}
with $\beta_k := (2k-1) / (2i-1)$. The associated eigenvector $\bm{v}_k$ of
$L_1$ is given by
\begin{equation}
  \label{e.v_rq}
  v_{kq} = b_k \cos \frac {\beta_k}{2} (2q-1)\pi,
  \quad (q=1,\dots,i-1).
\end{equation}
Here, $b_k$ is a constant, which can be obtained with the condition in
Eq.~(\ref{e.1Q=1}), or equivalently, $\sum_q v_{kq} = 1$. Thus, we have
\begin{equation}
  b_k = (-1)^{k+1} 2 \tan \frac {\beta_k \pi}{2}.
\end{equation}
Then, the diagonal elements of $C$, which are defined by $c_k = \bm{v}_k^2$ in
Sec.~\ref{s.derivation-gle}, can be determined as
\begin{equation}
  \label{e.c_k}
  c_k = \sum_{q=1}^{i-1} v_{kq}^2
  =\frac {1}{2} \sum_{q=1}^{2i-1} v_{kq}^2
  =(2i-1) \tan^2 \frac {\beta_k \pi}{2},
\end{equation}
where, in the second equality, we used $v_{ki}=0$ and $v_{kq}^2 = v_{k(2i-q)}^2$
for $q=1,\dots,i-1$.  Replacing $i$ with $i':=N-i+1$ and $\beta_k$ with
$\beta_k' := (2k-1)/(2i'-1)$ in Eqs.~(\ref{e.lambda_k})--(\ref{e.c_k}), we
obtain similar results for $L_2$.

Accordingly, the resistance kernel in Eq.~(\ref{e.mu(t)}) is given by
\begin{align}
  \label{e.rouse.mu(t)}
  \gamma \mu(t)
  =&
  \frac {4\kappa}{2i-1} \sum_{k=1}^{i-1}
  \cos^2 \frac {\beta_k\pi}{2} e^{-\frac {4\kappa t}{\gamma} \sin^2 \beta_k\pi/2}
  \notag\\[0.1cm]
  &+
  \frac {4\kappa}{2i'-1} \sum_{k=1}^{i'-1}
  \cos^2 \frac {\beta_k'\pi}{2} e^{-\frac {4\kappa t}{\gamma} \sin^2 \beta_k'\pi/2}.
\end{align}
If the tagged bead is the middle one, i.e., $i=i'$, the
Eq.~(\ref{e.rouse.mu(t)}) is consistent with a result presented in
Ref.~\cite{vandebroek17}. However, Eq.~(\ref{e.rouse.mu(t)}) is valid for
arbitrary $i$ ($1 \leq i \leq N$).

If $i, i' \gg 1$, modes with small $k \ll i, i'$ are dominant at long time
$t \gg \gamma/\kappa$, and therefore approximations
$\sin \beta_k \pi / 2 \approx \beta_k\pi/2$ and $\cos \beta_k \pi / 2 \approx 1$
are justified \cite{vandebroek17}. Let us denote this time scale as
$t_0 := \gamma/\kappa$. As a result, we obtain an expression of the memory
kernel for the continuous Rouse model studied in Ref.~\cite{panja10b}.
In particular, an integral approximation reveals power-law decay
\cite{panja10b,vandebroek17}
\begin{equation}
  \label{e.rouse.mu(t).middle}
  \mu(t) \approx 2 \left(\frac {1}{\pi t_0 t}\right)^{1/2},
\end{equation}
which is valid for $t_0 \ll t \ll t^{\ast}$ with $t^{\ast}$ being the longest
relaxation time; for example, if $i\sim i'$, then
$t^{\ast} = t_0(2i-1)^2/\pi^2$, and $\mu(t)$ decays exponentially at
$t \geq t^{\ast}$. Note that Eq.~(\ref{e.rouse.mu(t).middle}) is independent of
$i$, whereas $t^{\ast}$ depends on $i$. The equation
(\ref{e.rouse.mu(t).middle}) is consistent with a result for the middle bead
(i.e., $i=i'$) reported in Ref.~\cite{vandebroek17}.

If $i \gg 1$ and $i' \sim 1$, however, we obtain a different result, because the
second term in the right-hand side of Eq.~(\ref{e.rouse.mu(t)}) is negligible at
$t \gg t_0$. It follows that
\begin{equation}
  \label{e.rouse.mu(t).end}
  \mu(t) \approx \left(\frac {1}{\pi t_0 t}\right)^{1/2}.
\end{equation}
Thus, the resistance kernel $\mu(t)$ depends on the location of the tagged bead
$i$ in the Rouse chain.
The Laplace transforms of Eqs.~(\ref{e.rouse.mu(t).middle}) and
(\ref{e.rouse.mu(t).end}) yield
\begin{equation}
  \label{e.mu(s).rouse.integral-approx}
  1+\hat{\mu}(s) \approx
  \begin{cases}
    2\left(t_0 s\right)^{-1/2}  & (i,i'\sim \frac N2), \\[0.1cm]
    \left({t_0 s}\right)^{-1/2} & (i\sim 1 \,\,\text{or}\,\, i' \sim 1).
  \end{cases}
\end{equation}


\begin{figure}[t!]
  \centerline{\includegraphics[width=8.2cm]{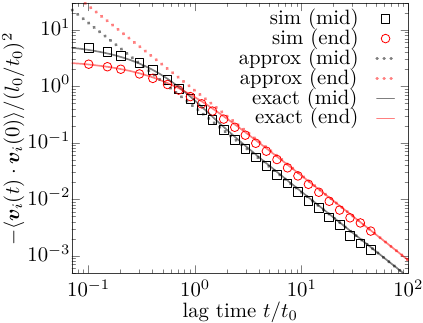}}
  \caption{\label{f.vcor} Velocity autocorrelation function
    $\left\langle \bm{v}_i(t)\cdot \bm{v}_i(0) \right\rangle$ for a tagged bead
    in the Rouse chain is presented with units of $t_0 := \gamma/\kappa$ and
    $l_0 = (k_BT/\kappa)^{1/2}$ (See Appendix \ref{app:nondim}). Because the
    velocity autocorrelation is negative, its absolute values are displayed in
    log-log form. In the simulation, the number of beads is set as $N=101$. The
    tagged bead is the middle bead ($i=51$; squares) and an end bead ($i=1$;
    circles). The solid lines represent the exact theory from
    Eq.~(\ref{e.rouse.phi(t)}), while the dotted lines represent the approximate
    theory from Eq.~(\ref{e.psi(t).rouse.integral-approx}).}
\end{figure}

\subsubsection {Mobility kernel}

Next, we derive the mobility kernel $\psi(t)$ for the Rouse model. The
eigenvalues and eigenvectors of $L$ in Eq.~(\ref{e.rouse.L}) are explicitly
given by \cite{amitai13}
\begin{align}
  \label{e.rouse.sigma_k}
  \sigma_k & = 4\kappa \sin^2 \frac {k\pi}{2N}, \\[0.1cm]
  \label{e.rouse.P_qk}
  p_{qk}   & = \sqrt{\frac {2- \delta_{k0}}{N}} \cos \frac {(2q-1)k\pi}{2N},
\end{align}
with $k=0,\dots,N-1$. Here, the eigenvectors are normalized as $\bm{p}_k^2=1$
[See Sec.~\ref{s.normal-mode.mobility}]. Inserting Eqs.~(\ref{e.rouse.sigma_k})
and (\ref{e.rouse.P_qk}) into Eq.~(\ref{e.psi.normal-mode}), we obtain the
mobility kernel for the Rouse model as
\begin{equation}
  \label{e.rouse.phi(t)}
  \psi(t) =
  - \frac {8\kappa}{N\gamma} \sum_{k=1}^{N-1}
  \cos^2 \frac {(2i-1)k \pi}{2N}
  \sin^2 \frac {k\pi}{2N}
  e^{-\frac {4\kappa t}{\gamma} \sin^2 \frac {k\pi}{2N}}.
\end{equation}

If $t \gg t_0$, then modes with small $k$ are dominant, and therefore the
approximation $\sin \pi k/2N \approx \pi k/2N$ can be justified.  Furthermore,
if $i \sim N/2$, the mobility kernel in Eq.~(\ref{e.rouse.phi(t)}) is
approximated as
\begin{equation}
  \label{e.psi(t).rouse.middle}
  \psi(t) \approx
  -\frac {\pi^2 \kappa}{N^3\gamma}
  \sum_{k=1}^{N-1} k^2 e^{-\left(\frac {k\pi}{N}\right)^2 \frac {\kappa}{\gamma} t}, 
\end{equation}
where we used $\cos^2 \pi(2i-1)k/2N \approx 1/2$. In contrast, if $i\sim 1$, we
obtain
\begin{equation}
  \label{e.psi(t).rouse.end}
  \psi(t) \approx
  -\frac {2\pi^2 \kappa}{N^3\gamma}
  \sum_{k=1}^{N-1} k^2 e^{-\left(\frac {k\pi}{N}\right)^2 \frac {\kappa}{\gamma} t}, 
\end{equation}
where we used $\cos \pi k/2N \approx 1$. This Equation
(\ref{e.psi(t).rouse.end}) coincides with a result for the Rouse model
reported in Ref.~\cite{sakaue13}, if we set a parameter $\tau_0$ employed in that
paper as $\tau_0 = \gamma/ \pi^2 \kappa$.

Integral approximations of Eqs.~(\ref{e.psi(t).rouse.middle}) and
(\ref{e.psi(t).rouse.end}) reveal power-law decay:
\begin{equation}
  \label{e.psi(t).rouse.integral-approx}
  \psi(t) \approx
  \begin{cases}
    -\frac {1}{4\sqrt{\pi}t_0}
    \left(\frac {t_0}{t}\right)^{3/2} &(i,i'\sim \frac N2),\\[0.1cm]
    -\frac {1}{2\sqrt{\pi}t_0}
    \left(\frac {t_0}{t}\right)^{3/2} &(i\sim 1 \,\,\text{or}\,\, i' \sim 1).
  \end{cases}
\end{equation}
The formulas in Eq.~(\ref{e.psi(t).rouse.integral-approx}) are valid for
$t_0 \ll t \ll t^{\ast}$, and correspond to the two formulas for the resistance
kernel $\psi(t)$ in Eqs.~(\ref{e.rouse.mu(t).middle}) and
(\ref{e.rouse.mu(t).end}). In fact, the Laplace transforms of
Eq.~(\ref{e.psi(t).rouse.integral-approx}) yield \cite{miyaguchi19}
\begin{equation}
  \label{e.psi(s).rouse.integral-approx}
  1+\hat{\psi}(s) \approx
  \begin{cases}
    \frac {1}{2} \left(t_0 s\right)^{1/2} & (i,i'\sim \frac N2), \\[0.1cm]
    \left(t_0 s\right)^{1/2}              & (i\sim 1 \,\,\text{or}\,\, i' \sim 1),
  \end{cases}
\end{equation}
where we used $\hat{\psi}(0) = -1$ from Eq.~(\ref{e.psi.1}).  By comparing
Eq.~(\ref{e.psi(s).rouse.integral-approx}) with
Eq.~(\ref{e.mu(s).rouse.integral-approx}), it is found that the relation in
Eq.~(\ref{e.pshi(t).mobility-kernel}) still holds under these approximations.

The mobility kernel is equivalent to the velocity autocorrelation if the
external force is absent [See Eq.~(\ref{e.<vi(t)vi(t')>})]. Therefore, from
Eq.~(\ref{e.psi(t).rouse.integral-approx}), it can be observed that the velocity
autocorrelation for the end beads is stronger than that for the middle
bead. This observation is consistent with the similar result for a polymerized
membrane \cite{keesman13}. The velocity autocorrelation functions for tagged
beads are numerically calculated and presented in Fig.~\ref{f.vcor}. The exact
formula in Eq.~(\ref{e.rouse.phi(t)}), displayed with solid lines in
Fig.~\ref{f.vcor}, is consistent with the numerical results across all time
scales, while the long-time approximations in
Eq.~(\ref{e.psi(t).rouse.integral-approx}), displayed with dotted lines, are
consistent with the numerical results at long times $t /t_0 > 1$.

Moreover, the power-law decay in the velocity autocorrelation $t^{-3/2}$ leads
to subdiffusion, characterized by the MSD $\sim t^{1/2}$ for the intermediate
time scale $t_0 \ll t \ll t^{\ast}$ \cite{panja10b}. At short and long times,
normal diffusion is observed, as generally demonstrated in
Eq.~(\ref{s.diffusivity_at_short_long}).

\subsection {Ring polymer}\label{s.ring-polymer}
\subsubsection {Resistance kernel}

The interaction matrix $L$ in Eq.~(\ref{e.rGNM.svec}) for the ring polymer is
given by
\begin{equation}
  \label{e.ring-polymer.L}
  L
  =
  \kappa
  \begin{pmatrix*}[r]
    2     & -1 & 0  & 0  &        &    & 0  & -1 \\
    -1    & 2  & -1 & 0  &        &              \\
    0     & -1 & 2  & -1 &        &              \\
          &    &    &    & \ddots &              \\
          &    &    &    &        & -1 & 2  & -1 \\
       -1 & 0  &    &    &        & 0  & -1 & 2
  \end{pmatrix*}.
\end{equation}
For the ring polymer, all the beads are identical, and therefore we define the
matrix $L'$ by removing the $N$th row and column from $L$ as
\begin{equation}
  \label{e.ring-polymer.L'}
  L'=
  \kappa
  \begin{pmatrix*}[r]
    2  & -1 & 0  &         &              \\
    -1 & 2  & -1 &         &              \\
       &    &    &  \ddots &              \\
       &    &    &         & -1 & 2  & -1 \\
       &    &    &         & 0  & -1 & 2
  \end{pmatrix*}.
\end{equation}

The eigenvalues and eigenvectors of $L'$ can be obtained by solving recursion
relations. The eigenvalues of $L'$ is given by
\begin{equation}
  \label{e.ring-polymer.lambda_k}
  \lambda_k = 4\kappa \sin^2 \frac {\pi k}{2N},
  \quad (k=1,\dots,N-1).
\end{equation}
The associated eigenvector of $L$, $\bm{v}_k$, is given by
\begin{equation}
  \label{e.ring-polymer.v_rq}
  v_{kq} = b_k \sin \frac {\pi k q}{N},
  \quad (q=1,\dots,i-1).
\end{equation}
Here, $b_k$ is a constant, which can be obtained with the condition in
Eq.~(\ref{e.1Q=1}), or equivalently, $\sum_q v_{kq} = 1$. Thus, we have
\begin{equation}
  b_k =
  \begin{cases}
    \tan \frac {\pi k}{2N} & (k: \text{odd}), \\[0.1cm]
    \infty                 & (k: \text{even}), 
  \end{cases}
\end{equation}
where, for even $k$, we have $\sum_q v_{kq} = 0$, and therefore we formally set
as $b_k = \infty$; this results in $c_k = \infty$ in
Eq.~(\ref{e.ring-polymer.c_k}) below (See also Sec.~\ref{s.diag-L-prime} for a
general explanation). Then, the diagonal elements of $C$, which are defined by
$c_k = \bm{v}_k^2$ in Sec.~\ref{s.derivation-gle}, can be determined as
\begin{equation}
  \label{e.ring-polymer.c_k}
  c_k = \sum_{q=1}^{N-1} v_{kq}^2
  = \sum_{q=1}^{N} v_{kq}^2
  =
  \begin{cases}
    \frac {N}{2} \tan^2 \frac {\pi k}{2N} & (k: \text{odd}), \\[0.1cm]
    \infty                                & (k: \text{even}).  
  \end{cases}
\end{equation}

Accordingly, the resistance kernel in Eq.~(\ref{e.mu(t)}) is given by
\begin{equation}
  \label{e.ring-polymer.mu(t)}
  \gamma \mu(t)
  =
  \frac {8\kappa}{N} \sum_{k=1, \text{odd}}^{N-1}
  \cos^2 \frac {\pi k}{2N} e^{-\frac {4\kappa t}{\gamma} \sin^2 \pi k/2N}.
\end{equation}
At long time $t \gg t_0 = \gamma/\kappa$, modes with small $k \ll N$ are
dominant and therefore approximations $\sin \pi k/ 2N \approx \pi k/2N$ and
$\cos \pi k/ 2N \approx 1$ are justified.  An integral approximation reveals the
same power-law decay as the middle bead in the Rouse model
[Eq.~(\ref{e.rouse.mu(t).middle})]. 

\subsubsection {Mobility kernel}

Here, we derive the mobility kernel $\psi(t)$ for the ring polymer. The
eigenvalues and eigenvectors of the symmetric circulant matrix $L$ in
Eq.~(\ref{e.ring-polymer.L}) are well known as \cite{berlin-kac52}
\begin{align}
  \label{e.ring-polymer.sigma_k}
  \sigma_k & = 4\kappa \sin^2 \frac {k\pi}{N}, \\[0.1cm]
  \label{e.ring-polymer.P_qk}
  p_{qk}   & = \sqrt{\frac 2N} \cos \left(\frac {2\pi kq}{N} + \frac {\pi}{4}\right),
\end{align}
for $k=0,\dots,N-1$. Note that the eigenvalues are degenerate because
$\sigma_k = \sigma_{N-k}$; hence, the term $\pi/4$ in the eigenvectors is
necessary to ensure orthogonality. The eigenvectors are normalized as
$\bm{p}_k^2=1$.

Inserting Eqs.~(\ref{e.ring-polymer.sigma_k}) and (\ref{e.ring-polymer.P_qk})
into Eq.~(\ref{e.psi.normal-mode}), we obtain the mobility kernel for the ring
polymer as
\begin{equation}
  \label{e.ring-polymer.phi(t)}
  \psi(t) =
  - \frac {4\kappa}{N\gamma} \sum_{k=1}^{N-1}
  \sin^2 \frac {k\pi}{N}
  e^{-\frac {4\kappa t}{\gamma} \sin^2 \frac {k\pi}{N}}.
\end{equation}
For $t \gg t_0$, modes with small $k \sim 1$ and large $k\sim N$ are dominant.
Therefore, using the approximation $\sin \pi k/N \approx \pi k/N$, we should
also double the formula.  Thus, we obtain
\begin{equation}
  \label{e.psi(t).ring-polymer.approx}
  \psi(t) \approx
  -\frac {8\pi^2 \kappa}{N^3\gamma}
  \sum_{k=1}^{N-1} k^2 e^{-\left(\frac {2k\pi}{N}\right)^2 \frac {\kappa}{\gamma} t}. 
\end{equation}
An integral approximation of Eq.~(\ref{e.psi(t).ring-polymer.approx}) reveals
exactly the same power-law decay as the middle bead of the Rouse model [the
first expression in Eq.~(\ref{e.psi(t).rouse.integral-approx})].


\section {Hydrodynamic interaction}\label{s.HI}

In this section, the GLE with the mobility kernel is derived for the elastic
network with the HI [Eq.~(\ref{e.rGNM.svec.HI})]; however, it is shown that the
GLE with the resistance kernel cannot be derived using the method in
Sec.~\ref{s.derivation_gle_r}.  We consider Eq.~(\ref{e.rGNM.svec.HI}) with the
external force $\bm{f}_{\mathrm{ex}}(t)$ applied to the tagged bead $i$:
\begin{equation}
  \label{e.rGNM.svec.HI.with.fex}
  \frac {d \bm{r}(t)}{dt} 
  =
  - L_H \cdot \bm{r}(t)  + \bm{h}_i\bm{f}_{\mathrm{ex}}(t) +  \bm{\xi}_H(t),
\end{equation}
where $\bm{f}_{\mathrm{ex}}(t)$ is a three-dimensional vector, and $\bm{h}_i$ is
the $i$th column vector of the mobility matrix $H$.


\subsection {GLE with two kernels}\label{s.derivation_gle_r_HI}

Following the derivations of Eqs.~(\ref{e.rGNM.svec.reduced}) and
(\ref{e.rGNM.tagged.reduced}) in Sec.~\ref{s.derivation_gle_r}, a generalization
to the case with the HI can be obtained from Eq.~(\ref{e.rGNM.svec.HI.with.fex})
as
\begin{align}
  \frac {d \bm{r}'(t)}{dt} 
  \label{e.rGNM.svec.reduced.HI}
  &=
  - L_H' \cdot [\bm{r}'(t) - \bm{r}_i(t) \bm{1}] +  \bm{\xi}_H'(t),
  \\[0.1cm]
  \frac {d \bm{r}_i(t)}{dt} 
  \label{e.rGNM.tagged.reduced.HI}
  &= \bm{\bar{w}}'\cdot L_H' \cdot [\bm{r}'(t) - \bm{r}_i(t) \bm{1}] + \bm{\xi}_{Hi}(t),
\end{align}
where Eq.~(\ref{e.(LHr)'}) is used, and we set $\bm{\bar{w}}':= \bm{w}/w_i$ and
$\bm{\xi}_{Hi} := (\bm{\xi}_H)_i$. Note that the law of action and reaction is
violated due to the HI in contrast to the case without HI
[Eqs.~(\ref{e.rGNM.svec.reduced}) and (\ref{e.rGNM.tagged.reduced})].

In exactly the same manner as the derivation of
Eq.~(\ref{e.gle.w.registance-kernel}), we obtain a GLE with mixed kernels,
\begin{align}
  &\frac {d \bm{r}_i(t)}{dt} + \int_0^t \mu_H(t-\tau) \dot{\bm{r}}_i(\tau) d\tau 
  \notag\\[0.1cm]
  \label{e.gle.w.registance-kernel.HI}
  &=\int_0^t \psi_H(t-\tau) \bm{f}_{\mathrm{ex}}(\tau) d\tau 
  +\bm{\xi}_{Hi}(t) + \bm{\xi}_{Hi}^{\mathrm{mix}}(t).
\end{align}
The resistance and mobility kernels in Eq.~(\ref{e.gle.w.registance-kernel.HI}),
$\mu_H(t)$ and $\psi_H(t)$, and the correlated noise
$\bm{\xi}_{Hi}^{\mathrm{mix}}(t)$ are defined by 
\begin{align}
  \label{e.mu.HI.1}
  \mu_H(t) &:= \bm{\bar{w}}' \cdot L_H' e^{- L_H' t} \cdot \bm{1},
  \\[0.1cm]
  \label{e.psi.HI.1}
  \psi_H(t) &:= \bm{\bar{w}}' \cdot L_H' e^{- L_H' t} \cdot \bm{h}_i',
  \\[0.1cm]
  \bm{\xi}_{Hi}^{\mathrm{mix}}(t) &:= \notag\\[0.1cm]
  \label{e.xi^r.HI.1}
  \bm{\bar{w}}' \cdot& L_H' \cdot
  \left[
  \int_0^t e^{- L_H'(t-\tau)} \cdot \bm{\xi}_H'(\tau)d\tau
  +
  e^{- L'_H t} \cdot \delta\bm{r}'(0)
  \right],
\end{align}
where $\bar{\bm{w}}' := \bm{w}'/ w_i$, and $\bm{h}_i'$ is the vector obtained by
removing the $i$th entry of $\bm{h}_i$.

Thus, we arrived at a GLE with two kernels, and it is not possible to derive the
GLE with only the resistance kernel using the method described in
Sec.~\ref{s.derivation_gle_r}. This is because
Eq.~(\ref{e.rGNM.svec.HI.with.fex}) is in a mobility representation. In other
words, to obtain the GLE with the resistance kernel, one would need to start
from a resistance representation; however, the method in
Sec.~\ref{s.derivation_gle_r} is not applicable in that case.

An exception is the case in which the beads have different masses and the HI is
absent. In this case, the mobility matrix $H$ is diagonal, and therefore
$\bm{h}_i' = \bm{0}$. Consequently, Eq.~(\ref{e.gle.w.registance-kernel.HI})
reduces to the GLE with the resistance kernel, satisfying the FDR
$\left\langle\bm{\xi}_{\mathrm{all}}(t)\bm{\xi}_{\mathrm{all}}(0)\right\rangle
=k_BT h_{ii}I_3 [\mu_H(t) + 2\delta(t)]$, where $h_{ii}$ is the $(i,i)$ entry of
$H$, and
$\bm{\xi}_{\mathrm{all}} := \bm{\xi}_{Hi}+\bm{\xi}_{Hi}^{\mathrm{mix}}$.

\subsection {GLE with mobility kernel}\label{s.gle-with-m-kernel_HI}

In contrast to the resistance representation in the previous subsection, the GLE
with a mobility representation can be derived from
Eq.~(\ref{e.rGNM.svec.HI.with.fex}) using the method described in
Sec.~\ref{s.gle-with-m-kernel} as
\begin{equation}
  \label{e.gle.w.mobility-kernel.HI}
  \frac {d\bm{r}_i(t)}{dt}
  =
  h_{ii}\bm{f}_{\mathrm{ex}}(t)
  + \int_0^t \psi_H(t-\tau) \bm{f}_{\mathrm{ex}}(\tau) d\tau
  + \bm{\xi}_{Hi} + \bm{\xi}_{Hi}^{\mathrm{m}}, 
\end{equation}
where the mobility kernel $\psi_H(t)$ and the correlated noise
$\bm{\xi}_{Hi}^{\mathrm{m}}(t)$ are defined as
\begin{align}
  \label{e.psi.1.HI}
  \psi_H(t) &:= -\bm{e}_i \cdot L_H e^{-L_H t} \cdot \bm{h}_i,
  \\[0.1cm]
  \label{e.xi^m.1.HI}
  \bm{\xi}_{Hi}^{\mathrm{m}}(t) &:= \notag\\[0.1cm]
  - \bm{e}_i &\cdot L_H \cdot
  \left[
  \int_0^te^{-L_H (t-\tau)} \cdot \bm{\xi}_H(\tau)d\tau
  +
  e^{-L_Ht} \cdot \delta\bm{r}(0)
  \right].
\end{align}
Thus, the HI affects the relaxation property of the tagged bead. If we define
$\bm{\xi}_{\mathrm{all}} := \bm{\xi}_{Hi}+\bm{\xi}_{Hi}^{\mathrm{m}}$, then the
FDR is given by
$\left\langle\bm{\xi}_{\mathrm{all}}(t)\bm{\xi}_{\mathrm{all}}(0)\right\rangle
=k_BTI_3 [\psi_{H}(t) + 2h_{ii} \delta(t)]$.


\section {Discussion}\label{s.discussion}

Previous studies have derived two types of the GLEs---one with the resistance
kernel and the other with the mobility kernel---for linear polymer models by
using normal modes \cite{panja10, panja10b, lizana10, sakaue13, saito15,
  vandebroek17,tian22}.  In this paper, these GLEs are derived for a broader
class of models: the linear elastic networks.  Notably, our derivation does not
rely on normal modes, and this makes the derivation process much clearer. We
also show that the two GLEs are interconnected through Laplace transform.

These derivations have been made possible thanks to the supervector notation
introduced in Sec.~\ref{s.reduced-vec-mat}. This notation is so useful that it
may be possible to derive a GLE for the case in which two beads are
simultaneously tagged.  To derive such a GLE, we should remove two rows and two
columns from $L$ ($i$th and $j$th rows and columns, say).  It is anticipated
that an interaction between tagged beads, mediated by the other beads, might
yield an additional coupling. Such a theoretical study should be crucial for
interpreting data from particle-tracking experiments, especially as recent
advances in experimental techniques enable simultaneously tracking multiple
points within a chromatin domain \cite{nozaki23}. Furthermore, for understanding
the fluctuations of distances between beads, such as the end-to-end distance
\cite{kou04, yamamoto14b, tian22}, the GLE with two tagged beads would become a
useful tool.

In general elastic networks, the equilibrium distance
$d_{ij}:= |\bm{r}_i - \bm{r}_j|$ between two beads at $T=0$ is non-zero
\cite{tirion96,thorpe07}; contrastingly, in the simple elastic network discussed
in this article, the distance $d_{ij}$ is assumed to be zero. Thus, in general
elastic networks, the matrix $LI_3$ should be replaced with a $3N\times 3N$
matrix $\tilde{L}$. As a result, in addition to the zero modes (eigenvectors
with the zero eigenvalue) due to translational symmetry, $\tilde{L}$ also has
zero modes due to rotational symmetry. Therefore, the matrix $\tilde{L}'$,
obtained by removing the tagged particle coordinates $\bm{r}_i$, is not positive
definite and is singular. Consequently, the derivation of the GLE with the
resistance kernel in Sec.~\ref{s.derivation_gle_r} must be modified;
specifically, a pseudo-inverse similar to $M$ in Eq.~(\ref{e.LML}) should be
employed.

If the external force is absent, the GLE with the mobility kernel in
Eq.~(\ref{e.gle.w.mobility-kernel}) becomes a simple Langevin equation with a
correlated noise
\begin{equation}
  \label{e.gle_m_wo_fex}
  \gamma \dot{\bm{r}_i}(t) = \bm{\xi}_i(t)+ \bm{\xi}_i^{\mathrm{m}}(t),
\end{equation}
where the two noise terms are mutually dependent. While equations of this form
without the white noise term $\bm{\xi}_i(t)$ have been frequently used
\cite{deng09, kursawe13, janczura21}, the mutual dependence of the two noise
terms has not been previously considered.  Therefore, this type of equation with
the complicated noise dependency should be explored in future
studies. Furthermore, it is necessary to determine whether the mutual dependency
of the noises is a general property of the GLE with the mobility kernel.

In addition, another GLE is derived by using a projection operator scheme, and
it is shown that the third GLE is consistent with the GLE with the resistance
kernel, although there is a subtle difference between the noise terms in the two
GLEs.  It prompts the question of whether there exists a projection operator
scheme consistent with the GLE with the mobility kernel, whose derivation is
considerably simpler than that of the GLE with the resistance kernel.

As important examples, the single bead dynamics in the Rouse model and the ring
polymer are elucidated using the general framework presented in
Sec.~\ref{s.normal-modes}. In particular, the resistance and mobility kernels
for the Rouse model depend on the position $i$ of the tagged bead in the Rouse
chain; for example, at long times, the antipersistence in velocity
autocorrelation for the end beads is stronger than that for the middle bead
[Eq.~(\ref{e.psi(t).rouse.integral-approx})]. It should be possible to analyze
simple polymer models with $L$ being a symmetric circulant matrix and
polymerized membranes \cite{keesman13,mizuochi14}.

A spectral density of states $g(\sigma)$ is the density of the eigenvalues of
the Kirchhoff matrix $L$ \cite{burioni04}. For instance, according to
Eq.~(\ref{e.rouse.sigma_k}), $g(\sigma)$ for the Rouse model is given by
$g(\sigma) \sim \sigma^{-1/2}$ as $\sigma \to 0$. Because $p_{ij}^2 \sim 1$ for
small $j$ for the Rouse model, the mobility kernel in
Eq.~(\ref{e.psi.normal-mode}) can be expressed as
$\gamma \psi(t) \sim - \int g(\sigma)\sigma e^{-\sigma t/\gamma} d\sigma \sim
-t^{-3/2}$ in consistent with Eq.~(\ref{e.psi(t).rouse.integral-approx}).
Interestingly, various folded proteins have been modeled by the elastic network
model, and the spectral density of states for these proteins has been estimated
as $g(\sigma)\sim \sigma^{\alpha}$ with $\alpha \in (0.4,1.1)$ \cite{burioni04}.
Therefore, assuming $p_{ij}^2 \sim 1$ and $\alpha \in (0,1)$, we obtain
$\gamma \psi(t) \sim - t^{-2-\alpha}$. Consequently, the MSD is given by
$\left\langle \bm{r}^2(t) \right\rangle\sim C_1 - C_\alpha t^{-\alpha}$, where
$C_1$ and $C_{\alpha}$ are positive constants. Thus, the MSD of folded proteins
would display power-law relaxation to a plateau at intermediate
timescales. These predictions should be verified in future studies (Observing
single monomer motion in real proteins is challenging; however, it may be
possible in molecular dynamics simulations).

If the interaction matrix $L_H$ can be estimated from experimental or numerical
data, it is possible to construct a GLE with a memory kernel, which is
determined by $L_H$ [Eq.~(\ref{e.psi.1.HI})]. This enables quantitative
coarse-grained modeling of complex molecules such as chromatin and proteins. In
fact, the estimation of $L$ has been attempted for chromatin using Hi-C data
\cite{shinkai20,shinkai22} and for proteins using data of their native
structures \cite{burioni04}. From the obtained $L$, the pre-averaged mobility
matrix $H$ can be determined analytically or numerically
\cite[Sec.~4.2]{doi86}. Thus, we could obtain $L_H = HL$ from experimental data,
and this coarse-grained model can be further validated using
single-particle-tracking data. Indeed, if trajectory data of single beads in
molecules are available from experiments or molecular dynamics simulations, the
MSD $\left\langle \bm{r}^2(t) \right\rangle$ can be calculated
\cite{uneyama15}. From the power-law behavior of the MSD at intermediate
timescales, the spectral density of states $g(\sigma)$ can be estimated. This
information, $g(\sigma)$, should be consistent with the eigenvalue spectrum of
the matrix $L_H$ estimated with the structure data.

Hydrodynamic interactions and excluded volume effects are important for
describing real polymer systems \cite{doi86}. The former was incorporated into
the present theory using a preaveraging approximation. For the Rouse model, the
pre-averaged HI can be further analyzed with normal modes \cite{sakaue13}, as
the preaveraged system is nearly diagonal in the normal modes of the Rouse model
without HI \cite{doi86}. However, it is unclear whether the preaveraged system
for the linear elastic model is also nearly diagonal in the normal modes of the
original system without HI. Moreover, the excluded volume effects could be
incorporated into the present model through a linearization approximation
\cite{doi86}, in which the Kirchhoff matrix $L$ (or $L_H$) should be modified to
be consistent with the equilibrium distribution $P_{\mathrm{eq}}(\bm{r})$.

Furthermore, recent single-particle-tracking experiments in cells have shown
that the diffusion coefficient of tagged particles appears to be random
\cite{parry14, lampo17, sabri20}. Hydrodynamic interaction is suggested as one
of the potential causes of such random diffusivity \cite{miyaguchi17,
  yamamoto21}. However, the preaveraging approximation completely disregards the
random fluctuations in diffusivity. Thus, improving the preaveraging
approximation to accurately capture the randomly fluctuating diffusivity is
crucial for a detailed understanding of experimental data. In addition, the
fluctuating diffusivity does not alter the behavior of the mean square
displacement, but it does affect higher-order moments such as the non-Gaussian
parameter \cite{uneyama15, miyaguchi19, uneyama19}.  Therefore, developing data
analysis methods to efficiently extract such higher-order information from time
series is necessary.

The normal-mode equations of motion in Eqs.~(\ref{e.dy/dt}) and
(\ref{e.dri/dt.normal-mode}) are equivalent to the equations derived in
Ref.~\cite[Eqs.~(13) and (14)]{miyaguchi22} from the GLE with the resistance
kernel by using a Markovian embedding method. These equations can be utilized
for numerical integration of the GLE. Similarly, the normal-mode equations in
Eq.~(\ref{e.dz/dt}) can serve as a numerical integration scheme for the GLE with
the mobility kernel. They should also provide a means to incorporate fluctuating
diffusivity into the GLE in Eq.~(\ref{e.gle_m_wo_fex}). Incorporating
fluctuating diffusivity into such GLEs has been a subject of recent research
\cite{wang20, wang20b, sabri20, janczura21, dieball22}, and the normal-mode
equations in Eq.~(\ref{e.dz/dt}) offer another approach to achieve this.

In the present paper, only linear models are studied. It would be important for
applications to investigate how nonlinear potential forces alter the GLEs
\cite{milster24}. In particular, incorporation of nonlinearity should make the
stochastic process non-Gaussian. It might be interesting to attempt deriving a
GLE for systems with weak nonlinearity using some perturbation method. If the
nonlinearity is sufficiently weak, it should be possible to analytically derive
a GLE using the projection operator method \cite{dhont96}.  Furthermore,
nonequilibrium dynamics due to the external force $\bm{f}_{\mathrm{ex}}(t)$
should also be investigated in future studies \cite{sakaue13, saito15,
  bingyu18}.

\begin{acknowledgments}
  We thank Dr. Takuma Akimoto and Dr. Eiji Yamamoto for fruitful discussion.
  S.S. was supported by JSPS KAKENHI Grant No. JP23H04297 and JST CREST Grant
  No. JPMJCR23N3. T.M. was supported by JSPS KAKENHI Grant No. JP22K03436.
\end{acknowledgments}

\appendix {}
\section {Positive definiteness of $L'$}\label{s.positiv-definite}

First, let us assume that $L$ has a zero eigenvalue and $N-1$ positive
eigenvalues. The eigenvector corresponding to the zero eigenvalue is
$\bm{1}_N = (1,\dots,1)$. We define an $N\times N$ matrix $L^{(i)}$ as follows.
The $i$th row and column of $L^{(i)}$ are zero, and the other elements are the
same as those of $L$.  Then, $L$ and $L^{(i)}$ are congruent; $S^tLS= L^{(i)}$
with a matrix
$S := [\bm{e}_1, \bm{e}_2, \dots, \overset{(i)}{\bm{1}}_{\!N}, \dots,
\bm{e}_N]$, where $\bm{e}_j$ is the $N$-dimensional unit vector
$\bm{e}_j := (0,\dots,\overset{(j)}{1},\dots,0)$.  Thus, due to the Sylevester's
law of inertia \cite[Sec.~4.5]{horn12book}, $L^{(i)}$ has a zero eigenvalue and
$N-1$ positive eigenvalues. It is then evident that the non-zero eigenvalues of
the $L^{(i)}$ are equivalent to the eigenvalues of $L'$. It follows that $L'$ is
positive definite.

\section {Proof of $AJ^tJB = AB$}\label{s.AJ2B=AB}

The matrix $J$ defined in Sec.~\ref{s.gle-with-m-kernel} is explicitly given by
\begin{equation}
  J = [\bm{e}_1,\dots, \bm{e}_{i-1}, \bm{e}_{N}, \bm{e}_{i},\dots, \bm{e}_{N-1}].
\end{equation}
Let $\bm{a}_j$ be the $j$th
column vector of $A$, and $\bm{b}_j$ be $j$th row vector of $B$. Then, we have
\begin{align}
  AJ^t & = [\bm{a}_1,\dots, \bm{a}_{i-1}, \bm{a}_{i+1},\dots, \bm{a}_{N}, \bm{a}_i],
  \\[0.1cm]
  JB   & = [\bm{b}_1,\dots, \bm{b}_{i-1}, \bm{b}_{i+1},\dots, \bm{b}_{N}, \bm{b}_i]^t.
\end{align}
It follows that $AJ^tJB = \sum_{j=1}^{N} \bm{a}_j\bm{b}_j = AB$, where each
summand $\bm{a}_j\bm{b}_j$ is an $N \times N$ matrix
\cite[Sec.~0.2.6]{horn12book}.

\section {Numerical scheme}\label{app:nondim}

For numerical integration of the Rouse model in Sec.~\ref{s.rouse-model}, the
equations of motion are discretized as
\begin{equation}
  \label{e.app.dr/dt}
  \gamma \left[\bm{r}(t+dt) - \bm{r}(t)\right]
  =
  -\kappa \tilde{L}\cdot \bm{r}(t) dt
  + \sqrt{2 \gamma k_BT dt}\tilde{\bm{\xi}}(t),
\end{equation}
where $dt$ is the time step size, $\tilde{L}$ is defined as
$\tilde{L} := L/\kappa$ [See Eq.~(\ref{e.rouse.L})], and $\tilde{\bm{\xi}}$ is a
Gaussian noise with zero mean and unit variance. Note that $\tilde{L}$ and
$\tilde{\bm{\xi}}$ are dimensionless.  A discretized velocity $\bm{v}_i(t)$ for
the tagged particle is defined by
\begin{equation}
  \label{e.app.velocity}
  \bm{v}_i(t)
  := \frac {\bm{r}_i(t+dt) - \bm{r}_i(t)}{dt},  
\end{equation}
The velocity autocorrelation
$\left\langle \bm{v}_i(t) \cdot \bm{v}_i(0)\right\rangle$ is numerically
calculated with both time and ensemble averages.

The equation (\ref{e.app.dr/dt}) can be made dimensionless by using
transformations
\begin{align}
  \label{e.app.nondimensionalize}
  \tilde{\bm{r}}(\tilde{t}) &= \frac {\bm{r}(t)}{l_0}, \quad
  \tilde{t} = \frac {t}{t_0},
\end{align}
where $l_0 = (k_BT/\kappa)^{1/2}$ and $t_0 = \gamma/\kappa$.  In
Fig.~\ref{f.vcor}, theoretical and numerical results are presented with these
units.

Dimensionless equations of motion for the Rouse model were then given by
\begin{equation}
  \label{e.app.tilde.dr/dt}
  \tilde{\bm{r}}(\tilde{t}+d\tilde{t}) - \tilde{\bm{r}}(\tilde{t})
  =
  -\tilde{L}\cdot \tilde{\bm{r}}(\tilde{t}) d\tilde{t}
  + \sqrt{2 d\tilde{t}}\tilde{\bm{\xi}}(\tilde{t}).
\end{equation}
As a numerical scheme to integrate Eqs.~(\ref{e.app.tilde.dr/dt}), the Euler
method \cite{kloeden11} is used with step size $d\tilde{t} = 0.05$. It is
assumed that $\tilde{\bm{r}}(0)$ is in equilibrium; the equilibrium distribution
is implemented by letting
$d\tilde{\bm{r}}_m=\tilde{\bm{r}}_{m+1}(0) - \tilde{\bm{r}}_{m}(0)$ follow the
Gaussian distribution $\propto \exp(-d\tilde{\bm{r}}_m^2/2)$ \cite{doi86}.



%

\end {document}